 \documentclass[10pt,preprint]{aastex}  
 \usepackage{natbib}
\def\ang{\AA}

\def\gapprox{\lower.4ex\hbox{$\;\buildrel >\over{\scriptstyle\sim}\;$}}
\def\lapprox{\lower.4ex\hbox{$\;\buildrel <\over{\scriptstyle\sim}\;$}}

\shortauthors{ASCHWANDEN}
\shorttitle{Self-Organized Criticality}

\begin{document}

\title{Global Energetics of Solar Flares. IX. Refined Magnetic Modeling}

\author{Markus J. Aschwanden}

\affil{	Solar and Astrophysics Laboratory,
	Lockheed Martin Advanced Technology Center, 
        Dept. ADBS, Bldg.252, 3251 Hanover St., Palo Alto, CA 94304, USA; 
        (e-mail: \url{aschwanden@lmsal.com})}

\begin{abstract}
A more accurate analytical solution of the {\sl vertical-current approximation 
nonlinear force-free field (VCA3-NLFFF)} model is presented that includes 
besides the radial $(B_r)$ and the azimuthal $(B_\varphi)$ magnetic field
components a poloidal component $(B_{\theta} \neq 0)$ also. This new analytical
solution is of second-order accuracy in the divergence-freeness condition, and
of third-order accuracy in the force-freeness condition. We re-analyze the
sample of 173 GOES M- and X-class flares observed with the {\sl Atmospheric
Imaging Assembly (AIA)} and {\sl Helioseismic and Magnetic Imager (HMI)}
onboard the {\sl Solar Dynamics Observatory (SDO)}. The new code reproduces
helically twisted loops with a low winding number below the kink instability 
consistently, avoiding unstable, highly-twisted structures of the Gold-Hoyle 
flux rope type. The magnetic energies agree within $E_{VCA3}/E_W=0.99\pm0.21$
with the Wiegelmann (W-NLFFF) code. The time evolution of the magnetic field
reveals multiple, intermittent energy build-up and releases in most flares,
contradicting both the Rosner-Vaiana model (with gradual energy storage in the
corona) and the principle of time scale separation ($\tau_{flare} \ll \tau_{storage}$)
postulated in self-organized criticality models. The mean dissipated flare energy is 
found to amount to $7\%\pm3\%$ of the potential energy, or $60\%\pm26\%$ of the 
free energy, a result that can be used for predicting flare magnitudes based on 
the potential field of active regions.  
\end{abstract}


\section{		INTRODUCTION  				}

How can we improve the accuracy of magnetic field measurements and
the values of dissipated magnetic energies during solar flares ?
The {\sl potential minimum-energy theorem} (Priest 2014; p.117) 
states that the potential field represents the lowest energy state of 
the magnetic field in the solar corona, while the free energy (defined
by the difference between the potential field and the non-potential
field) is the only magnetic field component that can be dissipated
during solar flares and other coronal heating episodes. The free 
energy in active regions typically amounts to a fraction of 
$E_{free}/E_p \approx 1\%-25\%$ of the potential field energy 
(Aschwanden et al.~2014a; Paper I). In the past, instead of calculating 
the free energy, the potential energy has been used to estimate
the dissipated energy in solar flares (e.g., Emslie et al.~2012),
which led to over-estimates of flare energies up to two orders
of magnitude. Comparisons and relative magnitudes of the potential
field energy $E_p$, the non-potential energy $E_{np}$, the
free energy $E_{free}=(E_{np}-E_p)$, and the energy dissipated
in flares $E_{diss}$ (which can be deduced from the step-wise decrease 
of the free energy during a flare), have been calculated for a sample of 
173 M and X-class flares in Paper I (Fig.~13 therein). These plots
reveal that the dissipated flare energies are of the same magnitude
as the free energies, but are far below the potential or nonpotential
energy. However, many cases reveal dissipated flare energies in excess 
of the free energy, which is obviously unphysical, being a side-effect 
of an inaccurate measurement method, which is corrected in this study.

More generally, we can ask which magnetic field extrapolation model is
most suitable and most accurate to measure the free energy and its
evolution during flares. Traditional {\sl non-linear force-free field
(NLFFF)} codes have been found to produce large uncertainties in the
horizontal (transverse to the line-of-sight) magnetic field components 
(e.g., Wheatland et al.~2000; Wiegelmann 2004; Wiegelmann et al.~2006; 
2012), mostly due to the fact that the force-free magnetic field is 
extrapolated from photospheric magnetograms, although the photosphere 
is not force-free (Metcalf et al.~1995). Improvements have been attempted 
by preprocessing of the line-of-sight magnetograms by additional 
constraints that minimize the force-freeness and net torque balance 
(Wiegelmann et al.~2012), by applying a {\sl magneto-hydrostatic (MHS)} model 
(Zhu et al.~2013; Wiegelmann et al.~2017; Zhu and Wiegelmann 2018),
or different magnetic helicity computation methods (Thalmann et al.~2019). 
The magnitude of free energy in braided and twisted magnetic fields may 
also strongly depend on the spatial resolution (Thalmann et al.~2014)
and the temporal cadence (Sun et al.~2012, 2017). 

A novel method that 
circumvents this problem is the {\sl Vertical-Current Approximation (VCA)} 
NLFFF code, which performs automated tracing of coronal loops that are used 
to minimize the magnetic field direction differences between the
observed coronal loops (where the magnetic field is supposedly
force-free) and the fitted VCA-NLFFF solutions (Aschwanden 2013a).
The suitability of the VCA-NLFFF code has been proven by the fact
that a step-wise decrease of the free energy in M and X-class flares
has been detected in many analyzed flares (Paper I).
Examples of the time evolution of the potential energy and the free
energy during flares are shown in Figures 11 and 12 in Paper I,
with a comparison of the NLFFF and the VCA-NLFFF calculations. 
Those examples confirm the compatibility of the potential energy 
in both the standard Wiegelmann NLFFF code and the VCA-NLFFF code.
A significant stepwise decrease of the free energy is noticable 
in many cases of both codes, but short-term positive increases 
occur also, {\bf either due to loop twisting, a coronal illumination
effect, or new emerging (current-carrying) flux.} 
It appears to be a
property detected by the VCA-NLFFF code, since it does occur less pronounced
in the Wiegelmann NLFFF code {\bf with pre-processed data}. 
This raises the question what conditions
affect the accuracy of reliable measurements of the free energy
during flares? {\bf Are there additional variability effects in the
time evolution of the free energy ?}
Is the approximation of the VCA-NLFFF code not 
sufficiently accurate? We will be able to answer these questions 
with a new version of the VCA3-NLFFF code that is based on a more 
accurate analytical solution. Also, the data suggest a new
interpretation in terms of impulsive and intermittent vertical 
current injections, rather than long-term coronal storage of 
currents in helically twisted loops.

The content of this paper includes a more accurate analytical
solution of the non-linear force free field code (VCA3-NLFFF)
(Section 2), updates of the numerical code (Section 3),
data analysis and results of improved magnetic energies 
(Section 4), a discussion of magnetic energy issues (Section 5),
and conclusions (Section 6). 

\section{		MAGNETIC FIELD MODEL			}

Previously we developed a nonlinear force-free field code that
models the 3-D coronal magnetic field in (flaring or non-flaring)
active regions, which we call the ``old'' Vertical-Current Approximation
(VCA-NLFFF) code here (Aschwanden and Sandman 2010; Sandman et al.~2009;
Sandman and Aschwanden 2011; 
Aschwanden et al.~2012, 2014a, 2014b, 2015b, 2016a, 2016c; 2018
Aschwanden 2013a, 2013b, 2013c, 2015, 2016b, 2019;
Aschwanden and Malanushenko 2013; Warren et al.~2018).
It has an advantage over other NLFFF codes by
including the magnetic field of automatically traced
coronal loops, while standard NLFFF codes use the transverse
field component in the non-forcefree photosphere. The inclusion
of coronal field directions is accomplished by automated tracing
of 2D-projected loop structures in EUV images, such as from AIA/SDO.
In the following we describe the low-energy limit of the potential field, 
which is identical to earlier versions of the VCA-NLFFF code, and then 
provide a new, more accurate analytical solution for the nonpotential field,
which we call the ``new'' Vertical-Current Approximation Version 3
(VCA3-NLFFF) code here.

\subsection{	Potential Field Parameterization	}

The conceptually simplest representation of a magnetic potential field
that fulfills Maxwell's divergence-free (solenoidal)
condition ($\nabla \cdot {\bf B}=0$)
is a unipolar magnetic charge $j$ that is buried below the solar surface,
which predicts a spherically symmetric magnetic field
${\bf B}_j({\bf x})$ that points away from the buried unipolar charge,
and its field strength falls off with the square of the distance $r_j$,
\begin{equation}
        {\bf B_j}({\bf x})
        = B_j \left({d_j \over r_j}\right)^2 {{\bf r}_j \over r_j} \ ,
\end{equation}
where $B_j$ is the magnetic field strength at the solar surface 
above a buried magnetic charge, $(x_j, y_j, z_j)$ is the
subphotospheric position of the buried charge, $d_j$ is the depth of 
the magnetic charge, 
\begin{equation}
	d_j = 1-\sqrt{x_j^2+y_j^2+z_j^2} \ ,
\end{equation}
and the vector ${\bf r}_j$ between an arbitrary location ${\bf x}=(x,y,z)$ 
in the solar corona (were we desire to calculate the magnetic field) and 
the location $(x_j, y_j, z_j)$ of the buried charge has a distance vector
${\bf r}_j$ is,   
\begin{equation}
	{\bf r}_j=[x-x_j, y-y_j, z-z_j] \ .
\end{equation}
Of course, the concept of a magnetic charge is only valid as a far-field
approximation, which avoids the complexity of the near-field details for
sake of mathematical convenience. We choose a Cartesian coordinate system
$(x,y,z)$ with the origin in the Sun center and we use units of solar
radii, with the direction of $z$ chosen along the line-of-sight from Earth
to Sun center. For a location near disk center ($x \ll 1, y \ll 1$), the
magnetic charge depth is $d_j \approx (1-z_j)$.
Thus, the distance $r_j$ from the magnetic charge is 
\begin{equation}
	r_j = \sqrt{(x-x_j)^2+(y-y_j)^2+(z-z_j)^2} \ .
\end{equation}
The absolute value of the magnetic field $B_j(r_j)$ is simply a function of
the radial distance $r_j$ (with $B_j$ and $d_j$ being constants for a
given magnetic charge),
\begin{equation}
	B(r_j) = B_j \left({d_j \over r_j}\right)^2 \ .
\end{equation}

In order to obtain the Cartesian coordinates
$(B_x, B_y, B_z)$ of the magnetic field vector ${\bf B}_j({\bf x})$, 
we can rewrite Eq.~(1) with Eqs.~(2-5) as,
\begin{equation}
	\begin{array}{ll}
		B_x(x,y,z) &= B_j \left({d_j / r_j}\right)^2 
			     (x-x_j) / r_j \\
		B_y(x,y,z) &= B_j \left({d_j / r_j}\right)^2 
			     (y-y_j) / r_j \\
		B_z(x,y,z) &= B_j \left({d_j / r_j}\right)^2 
			     (z-z_j) / r_j \\
	\end{array} \ .
\end{equation}

We proceed now from a single magnetic charge to an arbitrary number
$N_{\rm m}$ of magnetic charges and represent the general magnetic field with a
superposition of $N_{\rm m}$ buried magnetic charges, so that the potential field 
can be represented by the superposition of $N_{\rm m}$ fields ${\bf B}_j$ from 
each magnetic charge $j=1,...,N_{\rm m}$,
\begin{equation}
        {\bf B}({\bf x}) = \sum_{j=1}^{N_{\rm m}} {\bf B}_j({\bf x})
        = \sum_{j=1}^{N_{\rm m}}  B_j
        \left({d_j \over r_j}\right)^2 {{\bf r_j} \over r_j} \ .
\end{equation}
The radial vector ${\bf r}_j$, the magnetic charge location 
$(x_j, y_j, z_j)$, the intersection of the vertical symmetry axis with the
photosphere at $(x_0, y_0, z_0)$, and an arbitrary coronal location $(x, y, z)$ 
are depicted in Fig.~1. The potential field of a point charge
is solely defined by the radial vector component $B_r$ (Fig.~1), 
by setting $B_{\varphi}=0$ and $B_{\theta}=0$.

\subsection{	Nonlinear Force-Free Field Parameterization	}

In order to improve the accuracy of the old VCA-NLFFF code we
searched for more accurate analytical solutions of the equation system
for the divergence-free,
\begin{equation}
        \nabla \cdot {\bf B} = 0 \ ,
\end{equation}
and force-free magnetic field,
\begin{equation}
        \nabla \times {\bf B} = \alpha({\bf x}) {\bf B}  \ ,
\end{equation}
where the nonlinear force-free parameter $\alpha({\bf x})$ is
spatially varying, but remains constant along a given magnetic
field line.

We can derive a nonlinear force-free field solution by writing the
divergence-free condition (Eq.~8) and the force-free condition 
(Eq.~9) of a magnetic field vector ${\bf B} = (B_r, B_\theta, B_\varphi)$ 
in spherical coordinates $(r, \theta, \varphi)$. The 
location of the magnetic charge and the spherical symmetry axis are aligned 
with the vertical direction to the local solar surface. Eqs.~(8-9) expressed
in spherical coordinates are then, 
\begin{equation}
       (\nabla \cdot {\bf B}) = {1 \over r^2} {\partial \over \partial r}
	(r^2 B_r)  
	+ {1 \over r \sin{\theta}} {\partial \over \partial \theta}
	(B_\theta \sin{\theta}) 
	+ {1 \over r \sin{\theta}}
	{\partial B_\varphi \over \partial \varphi} = 0 \ ,
\end{equation}
\begin{equation}
	\left[ \nabla \times {\bf B} \right]_r =
 	{1 \over r \sin{\theta}}
 	\left[{\partial \over \partial \theta}
 	(B_\varphi \sin{\theta}) -
	{\partial B_\varphi \over \partial \varphi} \right]
	= \alpha B_r \ ,
\end{equation}
\begin{equation}
	\left[ \nabla \times {\bf B} \right]_\theta =
 	{1 \over r}
 	\left[{1 \over \sin{\theta}} {\partial B_r \over \partial \varphi}
	-{\partial \over \partial r} (r B_\varphi) \right]
	= \alpha B_\theta \ ,
\end{equation}
\begin{equation}
	\left[ \nabla \times {\bf B} \right]_\varphi =
 	{1 \over r}
 	\left[{\partial \over \partial r}( r B_\theta ) 
	-{\partial B_r \over \partial \theta} \right]
	= \alpha B_\varphi \ .
\end{equation}
For a simple approximative nonlinear force-free solution we 
require axi-symmetry with no azimuthal dependence 
($\partial / \partial \varphi = 0$).
This requirement simplifies Eqs.~(10-13) to, 
\begin{equation}
        {1 \over r^2} {\partial \over \partial r} (r^2 B_r) 
	+ {1 \over r \sin{\theta}} {\partial \over \partial \theta}
	(B_\theta \sin{\theta}) = 0 
	\ ,
\end{equation}
\begin{equation}
 	{1 \over r \sin{\theta}}
	{\partial \over \partial \theta}(B_\varphi \sin \theta) - \alpha B_r 
	= 0 \ ,
\end{equation}
\begin{equation}
 	-{1 \over r} {\partial \over \partial r} (r B_\varphi) 
	- \alpha B_\theta = 0 ,
\end{equation}
\begin{equation}
 	  {1 \over r} 
 	{\partial \over \partial r}( r B_\theta ) 
	- {1 \over r} {\partial B_r \over \partial \theta} 
	- \alpha B_\varphi = 0 \ .
\end{equation}

An approximate solution of this system of four differential equations
has been analytically derived for the radial field component
$B_r(r, \theta)$ (Eq.~18), the azimuthal field component
$B_{\varphi}(r, \theta)$ (Eq.~19), and the nonlinear 
parameter $\alpha(r, \theta)$ (Eq.~21), for the special case of
a vanishing poloidal field component $B_\theta (r, \theta)=0$
(Aschwanden 2013a). Here we find a new (approximative) analytical 
solution for the poloidal field component $B_\theta (r, \theta) \neq 0$,
as expressed in (Eq.~20),
\begin{equation}
	B_r(r, \theta) = B_0 \left({d^2 \over r^2}\right)
	{1 \over (1 + b^2 r^2 \sin^2{\theta})} \ , 
\end{equation}
\begin{equation}
	B_\varphi(r, \theta) = 
	B_0 \left({d^2 \over r^2}\right)
	{b r \sin{\theta} \over (1 + b^2 r^2 \sin^2{\theta})} \ ,
\end{equation}
\begin{equation}
	B_\theta(r, \theta) = 
	B_0 \left({d^2 \over r^2}\right)
	{b^2 r^2 \sin^3(\theta) \over
	(1 + b^2 r^2 \sin^2 \theta )} {1 \over \cos \theta}
	\ ,
\end{equation}
\begin{equation}
	\alpha(r, \theta) \approx {2 b \cos{\theta} \over 
	(1 + b^2 r^2 \sin^2{\theta})}  \ .
\end{equation}
{\bf where the constant $b$ is defined in terms of the number 
of full twisting turns $N_{twist}$ over the loop length $L$, 
i.e., $b = 2 \pi N_{twist}/L$.} 
In the old VCA-NLFFF code we neglected the poloidal magnetic
field component (i.e., $B_{\theta}=0$), because the magnitude of
this component is of second order $ \propto (b r \sin \theta)^2$
like the residuals of the VCA approximation. In the new VCA-NLFFF
code presented here, however,
we increase the accuracy of the nonpotential field
solutions by taking an approximate solution of $B_\theta \neq 0$ (Eq.~20)
into account. Inserting this new solution (Eqs.~18-21) into the
divergence-free term (Eq.~14) and the force-free terms (Eqs.~15-17),
we find that the solution is accurate
to second order in $\nabla \cdot {\bf B}$ (Eq.~10),
to third order in $[\nabla \times {\bf B}]_\varphi$ (Eq.~13),
to fifth order in $[\nabla \times {\bf B}]_\theta$ (Eq.~12),
and being exactly zero for the radial component
$[\nabla \times {\bf B}]_r = 0$ (Eq.~11). A proof of this
approximative analytical solution is provided in Appendix A. 

\subsection{	Cartesian Coordinate Transformation 		}

In the previous derivation we derived the solution in terms
of spherical coordinates $(r, \theta, \varphi)$ in a coordinate system
where the rotational symmetry axis is aligned with the vertical to the
solar surface intersecting a magnetic charge $j$ (Fig.~1), expressed
with the 3-D vectors $B_r(r, \theta)$, $B_{\varphi}(r, \theta)$, and
$B_\theta(r, \theta)$. 
For a model with multiple magnetic charges at arbitrary positions on the
solar disk, we have to transform the individual coordinate systems
$(r_j, \theta_j, \varphi_j)$ associated with magnetic charge $j$
into a Cartesian coordinate system $(x,y,z)$
that is given by the observers line-of-sight (in $z$-direction) and
the observer's image coordinate system $(x,y)$ in the plane-of-sky.
The geometric relationships of the spherical coordinate vectors
are shown in Fig.~1.

Defining the radial vectors ${\bf r}_j$ (between $(x_j, y_j, z_j)$
and $(x, y, z)$), and ${\bf R}$ between the solar center $(0, 0, 0)$
and a magnetic charge $(x_j, y_j, z_j)$, as well as the vector products 
between the orthogonal spherical components $(B_r, B_{\varphi}, B_\theta)$, 
we obtain the following directional cosines of three spherical 
vector components (see Fig.~1),
\begin{equation}
	{\bf r}_j=[x-x_j, y-y_j, z-z_j] \ .
\end{equation}
\begin{equation}
	{\bf R} = \left[ x_j, y_j, z_j \right] \ ,
\end{equation}
\begin{equation}
	{{\bf B}_{r} \over B_r} =  
	\left[ {x-x_j \over r_j}, {y-y_j \over r_j}, {z-z_j \over r_j}\right]
	= \left[ \cos_{r,x}, \cos_{r,y}, \cos_{r,z} \right] \ ,
\end{equation}
\begin{equation}
	  {{\bf B}_{\varphi} \over B_{\varphi}}
	= { {\bf R} \times {{\bf B}_r} \over |{\bf R} \times {\bf B}_r|} 
	= \left[ \cos_{\varphi,x}, \cos_{\varphi,y}, \cos_{\varphi,z} \right]\ ,
\end{equation}
\begin{equation}
	  {{\bf B}_{\theta} \over B_{\theta}}
	= { {\bf B}_r \times {{\bf B}_\varphi} \over |{\bf B}_r \times {\bf B}_\varphi|} 
	= \left[ \cos_{\theta,x}, \cos_{\theta,y}, \cos_{\theta,z} \right]\ ,
\end{equation}
The vector product allows us also to extract the inclination angle 
$\theta_j$ between the radial magnetic field component ${\bf B}_r$ and the 
local vertical direction ${\bf R}$,
\begin{equation}
	\theta_j = \sin^{-1} \left( { |{\bf R} \times {\bf B}_r|
	\over |{\bf R}| \ |{\bf B}_r| } \right) \ .
\end{equation}
The transformation of the non-potential field from spherical coordinates 
$(B_r, B_\varphi, B_\theta)$ into cartesian coordinates $(B_x, B_y, B_z)$,
which takes the sphericity of the solar surface fully into account,
amounts to 
\begin{equation}
	\begin{array}{ll}
	B_x &= B_r(r_j, \theta_j) \cos_{r,x} 
		+ B_{\varphi} (r_j, \theta_j) \cos_{\varphi, x} 
		+ B_\theta (r_j, \theta_j) \cos_{\theta, x} \\
	B_y &= B_r(r_j, \theta_j) \cos_{r,y} 
		+ B_{\varphi}(r_j, \theta_j) \cos_{\varphi, y} 
		+ B_\theta (r_j, \theta_j) \cos_{\theta, y} \\
	B_z &= B_r(r_j, \theta_j) \cos_{r,z} 
		+ B_{\varphi}(r_j, \theta_j) \cos_{\varphi, z} 
		+ B_\theta (r_j, \theta_j) \cos_{\theta, z} \\
	\end{array} \ ,
\end{equation}
This is a convenient parameterization that allows us directly to
calculate the magnetic field vector of the non-potential field 
${\bf B}_j=(B_x, B_y, B_z)$ associated with
a magnetic charge $j$ that is characterized with five parameters:
$(B_j, x_j, y_j, z_j, \alpha_j)$, where the force-free
$\alpha$-parameter is related to the twist parameter 
in terms of the number of twists $N_{twist}$ and the loop length $l$
by $b_j=2 \pi N_{twist}/l$.

\subsection{ 	Superposition of Twisted Field Components 	}

The total non-potential magnetic field from all $j=1,...,N_{\rm m}$ magnetic
charges can be approximately obtained from the vector sum of 
all components $j$ (in an analogous way as we applied in Eq.~(7) 
for the potential field), 
\begin{equation}
        {\bf B}({\bf x}) = \sum_{j=1}^{N_{\rm m}} {\bf B}_j({\bf x}) \ ,
\end{equation}
where the vector components ${\bf B}_j=(B_{x,j}, B_{y,j}, B_{z,j})$
of the non-potential field of a magnetic charge $j$ are defined in 
Eq.~(28), which can be parameterized with $5 N_{\rm m}$ free parameters 
$(B_j, x_j, y_j, z_j, \alpha_j)$ for a non-potential field, or with
$4 N_{\rm m}$ free parameters for a potential field (with $\alpha_j = 0$). 

Let us consider the condition of divergence-freeness.
Since the divergence operator is linear, the superposition of a number of
divergence-free fields is divergence-free also,
\begin{equation}
        \nabla \cdot {\bf B} = \nabla \cdot (\sum_j {\bf B}_j)
        = \sum_j (\nabla \cdot {\bf B}_j) = 0 \ .
\end{equation}

Now, let us consider the condition of force-freeness.
A force-free field has to satisfy Maxwell's equation (Eq.~9). 
Since we parameterized both the potential field and the non-potential
field with a linear sum of $N_{\rm m}$ magnetic charges, the requirement
would be, 
\begin{equation}
	\nabla \times {\bf B} =
	\nabla \times \sum_{j=1}^{N_{\rm m}} {\bf B}_j = 
	\sum_{j=1}^{N_{\rm m}} (\nabla_j \times {\bf B}_j) = 
	\sum_{j=1}^{N_{\rm m}} \alpha_j ({\bf r}) {\bf B}_j = 
	\alpha ({\bf r}) {\bf B} \ .
\end{equation}
Generally, these three equations of the vector ${\nabla \times \bf B}$
cannot be fulfilled with a scalar function $\alpha({\bf r})$ for a
sum of force-free field components, unless the magnetic field volume 
can be compartmentalized by spatially separated sub-volumes. 
Since the magnetic field of each point source decreases with the square 
of the distance, the force-free field solution of one particular 
magnetic charge is not much affected by the contributions from a 
spatially well-separated secondary magnetic charge, and thus Eq.~(31) 
is approximately valid for spatially well-separated sources. Our derivation
of the analytical solution of the magnetic field (shown in Appendix A
and summarized in Table 1) proves that the force-free condition 
(Eq.~31) is fulfilled with an accuracy of the third-order
term of $(b r \sin \theta)$. In practice, this means that the
force-free field at a location with $(b r \sin \theta) \le 0.1$
has an accuracy of $(b r \sin \theta) \le 0.001$. 

\subsection{	The Gold-Hoyle Flux Rope Test		} 

We demonstrate the improvement of the new VCA3-NLFFF code in various
ways. One test consists of calculating the figures of merit for the
divergence-freeness and the force-freeness, integrated over the
simulated computation box (Eqs.~48-49 in Aschwanden 2013a). 
Such tests yield very similar values for the old VCA-NLFFF and the
new VCA3-NLFFF code. A second test is the display of poloidal field lines, 
which should show a curvature in poloidal planes, as it is expected for
a correction from $B_\theta=0$ to $B_\theta \neq 0$, which is indeed 
the case and confirms the expected change in the geometry of magnetic field
lines (Fig.~2, panel b versus d).

A particularly instructive test is the reproduction of a Gold-Hoyle
flux rope (Gold and Hoyle 1960), which is thought to have a large 
number of multiple helical windings, as inferred from eruptive flux 
ropes propagating along with coronal mass ejections (CMEs) through 
the heliosphere (Amari et al.~1996; Gibson and Low 2000; Wang et al.~2017).
An attempt to reproduce such flux ropes with
1-5 full turns was carried out in Aschwanden (2013a; Appendix A),
which revealed large distortions from a uniformly thick flux tube
geometry (Fig.~3 left panels). With the new VCA3-NLFFF code we
find that the number of helical windings does not exceed more
than about one full turn (Fig.~3 right panels),
no matter how large the non-potential
force-free $\alpha$ parameter (or $b$) is chosen, which implies
that helical geometries with more than one full turn cannot be
reproduced with the new VCA3-NLFFF code. The most likely explanation
for this limit is that helical magnetic field lines with a twist
of more than one turn are not force-free, as it is theoretially
expected from the kink instability criterion (Priest 2014;
Hood and Priest 1979; T\"or\"ok et al.~2003; Kliem and T\"or\"ok 2006; 
Kliem et al.~2014), as well
as from observational evidence ($N_{twist} \lapprox 0.5$) of
the helical twisting number (or the Gauss braiding linkage number) of
coronal loops in solar active regions (Aschwanden 2019).
Helical flux ropes as envisioned by Gold and Hoyle (1960) may still
exist in interplanetary space, but are transient and require a 
time-dependent solution of the non-forcefree magnetic field.

\section{	NUMERICAL CODE 					}

Previously we developed the original {\sl vertical-current
approximation nonlinear force-free field code (VCA-NLFFF)} code 
that has been described and continuously improved over a decade
(Aschwanden and Sandman 2010; Sandman and 
Aschwanden 2011; Aschwanden et al.~2012, 2014a, 2016a;
Aschwanden 2013a, 2013b, 2013c, 2015, 2016b, 2019;
Aschwanden and Malanushenko 2013; Warren et al.~2018).
In the following we mention the most significant changes 
in the data analysis method of the new VCA3-NLFFF code only.

\subsection{	Improvements of the VCA3-NLFFF Code		}

Over the duration of the SDO mission (Pesnell at al.~2011)
from 2010 to 2019, some changes in the scaling of the EUV 
brightness of AIA images (from a factor of unity to a factor 
of 0.1) occurred, which are now automatically corrected
by the AIA and HMI data reading software, based on the
FITS descriptor value BSCALE.

For the absolute calibration of the magnetic energy we compare
the line-of-sight magnetic field component $B_z^{map}(x,y)$ 
of the observed HMI magnetogram with the modeled component
$B_z^{model}(x,y)$ obtained from fitting each local peak
in the magnetogram with the expected field from a buried
unipolar magnetic charge. We obtain equivalence of the
corresponding magnetic energy component for an empirical
correction factor of $q_B \approx 0.8$, i.e.,
\begin{equation}
	q_e^{model} = { \sum_{x,y} (q_B \times B_z^{model})^2
		  \over \sum_{x,y} (B_z^{map})^2 } \ ,
\end{equation}
where the relationship $E \propto B^2$ is used for the relationship
between the magnetic energy $E$ and the magnetic field strength $B$.
This correction factor of $q_B \approx 0.8$ accounts for
missing magnetic flux in the model that results from overlapping
magnetic field domains where oppositie magnetic polarities 
(positive and negative magnetic fields) cancel out parts of the field.
Note that a correction factor of $q_B=0.8$ of the average field
strength corresponds to a factor of $q_e \approx 0.8^2 = 0.64$ 
in magnetic energy. This correction propagates in the various forms 
of uncorrected magnetic energies ($E_p^*, E_{np}^*$) as 
\begin{equation}
	E_p(t) = E_p^{*}(t) / q_e^{model} \ ,
\end{equation}
\begin{equation}
	E_{np}(t) = E_{np}^{*}(t) / q_e^{model} \ ,
\end{equation}
\begin{equation}
	E_{free}(t) = E_{np}(t) - E_p(t) =
		     [E_{np}^{*}(t) - E_p^{*}]/q_e^{model} \ .
\end{equation}
This empirical correction supersedes a geometric twist correction
factor $q_{iso} \approx (\pi/2)^2 \approx 2.5$ used in previous work
(Eq.~2 in Paper I), which represents another approach to recover 
under-estimated free energy. 

{\bf A specialty of the VCA3-NLFFF code is the usage of EUV images
(from AIA/SDO), in addition to the line-of-sight magnetograms
(with $B_z(x,y)$ from HMI/SDO), in order to constrain the non-linear 
force-free field solution, rather than using the transverse
$(B_x, B_y)$ magnetic field components. We use AIA images in all 
6 coronal wavelengths. An automated pattern recognition code, which
is specialized to measure the projected 2-D coordinates $[x(s), y(s)]$
of curvi-linear features is then employed to extract coronal loop
segments (mostly in the lowest density scale height of the corona)
at altitudes of $2-50$ Mm (see yellow curves in Figs.~4 and 5). 
Several hundred loop features are extracted
in a set of 6 wavelength images, which are further subdivided into
7 positions per segment, and for each location the line-of-sight
coordinates $z(s)$ are optimized by varying the non-linear force free
$\alpha$-parameter for each unipolar magnetic charge (typically 30
values $\alpha_i, i=1,...,30$).}
 
The VCA3-NLFFF code converges to a best-fit solution of the average
misalignment angle between the observed magnetic field directions and
the theoretical magnetic field model for every time step {\bf (indicated
with $\mu_2$ in Figs.~4 and 5, found in a range of $\mu_2 \approx 
5.9^\circ-20.6^\circ$ here).} The numerical
convergence can be improved if the solution from a previous time
step $t_i$ is used as initial condition in the iterative convergence
of subsequent time steps $t_{i+1}$. The convergence behavior of our
VCA3-NLFFF code is found to be stable, but slightly oscillates between
two subsequent time steps. In order to take advantage of this numerical
behavior, we smooth the time evolutions of the energies by averaging
two subsequent solutions, which produces a smoother time evolution 
$E(t)$ in each of the numerically obtained energies (Eqs.~33-35),
\begin{equation}
	E(t) = {E(t_i) + E(t_{i+1}) \over 2} \ . 
\end{equation}

The most important correction of the improved VCA3-NLFFF code is the
introduction of an analytical expression for the poloidal magnetic field 
vector $B_{\theta} ( r, \theta )$ (Eq.~20), which was neglected in the 
original code, i.e., $B_\theta (r, \theta)=0)$, as well as the 
related coordinate transformations of the spherical magnetic field 
components of ${\bf B}_\theta (r, \theta)$ (Eq.~26) into cartesian coordinates
${\bf B}=(B_x, B_y, B_z)$, that depend on spherical coordinates
$(r, \theta, \varphi)$ as derived in (Eq.~28).

\section{	DATA ANALYSIS AND RESULTS			}

We analyzed the same data set of 173 solar flares presented in Paper I,
which includes all M- and X-class flares observed with the SDO
(Pesnell et al.~2011) during the first 3.5 years of the mission
(2010 June 1 to 2014 January 31). This selection of events has a
heliographic longitude range of $[-45^\circ, +45^\circ]$, for which
magnetic field modeling can be faciliated without too severe
foreshortening effects near the solar limb. We use the 45-s 
line-of-sight magnetograms from HMI/SDO and make use of all
coronal EUV channels of AIA/SDO (in the six wavelengths
94, 131, 171, 193, 211, 335 \ang ), which are sensitive to
strong iron lines (Fe VIII, IX, XII, XIV, XVI, XVIII, XXI, XXIV) 
in the temperature range of $T \approx 0.6-16$ MK. The spatial
resolution is $\approx1.6"$ (0.6" pixels) for AIA, and
the pixel size of HMI is 0.5". The coronal magnetic field is
modeled by using the line-of-sight
magnetogram $B_z(x,y)$ from HMI and (automatically detected)
projected loop coordinates $[x(s), y(s)]$ in each EUV wavelength of AIA.
A full 3-D magnetic field model ${\bf B}(x,y,z)$ is computed 
for each time interval and flare with a cadence of 6 min
(0.1 hrs), where the total duration of a flare is defined
by the GOES flare start and end times, including a margin
of 30 minutes before and after each flare. The size of
the computation box amounts to an area with a width and length
of 0.5 solar radii in the plane-of-sky, and an altitude range
of 0.2 solar radius. The total number of analyzed data 
includes 2706 HMI images and 16,236 AIA images.

\subsection{	Time Evolution of Magnetic Energies		}

The 3-D magnetic field solutions obtained with the VCA3-NLFFF code
are shown in form of magnetic field lines in Figs.~4 and 5, while
we present in Figs.~6 and 7 the time evolution of the magnetic potential 
energy $E_p(t)$ (orange diamonds), the magnetic nonpotential energy $E_{np}(t)$
(red diamonds), the free energy $E_{free}(t)$ (red diamonds), 
the dissipated magnetic energy
$E_{diss}$, and the time derivative (black hatched areas) of the GOES flux 
$F_{GOES}(t)$ (dashed curve), which is a proxy for the hard X-ray time profile. 
We compare the energies obtained with the VCA3-NLFFF code (red curves with
diamonds) with those from the Wiegelmann (W-NLFFF) code (blue curves with crosses).
The time evolution of the free energy $E_{free}(t)$ is computed from the
cumulative negative decreases (dashed red curves) in the time interval 
between the flare start time and end time (black dotted lines in Figs.~6 and 7).
The time intervals of negative decreases are also marked with red
hatched areas in Figs.~6 and 7. Note that they often coincide with the
peak time of the GOES time derivative (solid vertical black line in Figs.~6 and 7).
The uncertainties of the dissipated energies $E_{diss}$ were estimated from the
median values of energy decreases in subsequent time steps.

We see different degrees of agreements between the two codes.
For example, an excellent agreement between the VCA3-NLFFF and the W-NLFFF 
code is obtained for flare \#67 (Fig.~6d), which shows a dissipated energy
of $E_{diss}$[W-NLFFF]=$86 \times 10^{30}$ erg, and $E_{diss}$[VCA3-NLFFF]=
$(76 \pm 7) \times 10^{30}$ erg. In another example, the expected step function
of the free energy prominently shows up in the flare \#384 (Fig.~7e)
with a value of $E_{diss}$[VCA3-NLFFF]=$(639 \pm 31) 10^{30}$ erg, but the
W-NLFFF code yields a ten times smaller step with 
$E_{diss}$[W-NLFFF]=$62 \times  10^{30}$ erg. There is one case
(flare \#148, Fig.~6f) with no energy decrease detected with $E_{diss}$[W-NLFFF],
while $E_{diss}$[VCA3-NLFFF]=$(88 \pm 12)$ shows clearly a significant
magnetic energy decrease. So, there are obvious discrepancies between 
the W-NLFFF and the VCA3-NLFFF codes, but at this point we cannot
determine which code is more accurate. 

\subsection{	Complexity of Magnetic Field Structures 	}

We show the forward-fitted magnetic field solutions 
for the 11 X-class flares (out of the 173 M- and X-class events) in Figs.~4 and 5.
We see that major sunspots with opposite magnetic polarities
(white and black parts of the magnetograms in Figs.~4 and 5) occur in each
of the flaring active regions, but the interface between opposite
magnetic polarities can be quite complex. Examples of 
isolated leading sunspots are observed in the flares \#
66, 349, 351, and 384 (Figs.~4, 5), while the other cases
manifest mixed polarity sunspots, often with sharp gradients
in the magnetic field, such as in the flares \#12, 147, and 148
(Figs.~4, 5). These details are important, because the degree of complexity 
in mixed magnetic polarities determines how well the data are suited
to fit the theoretical model of buried unipolar magnetic charges.
The more isolated, isotropic, and axi-symmetric a sunspot is,
the better it can be fitted with a buried unipolar
charge, and a higher accuracy of the fitted magnetic field strengths
is expected. In other words, the sharper the magnetic field gradients
become between positive and negative magnetic polarities, the 
less accurate is the modeling with unipolar point charges. 
We will come back to this issue in the next section
on the accuracy of absolute magnetic field strengths.

\subsection{	Absolute Magnetic Field Comparison 	}

A first quantitative test of our study is how well the VCA3-NLFFF code can
retrieve the absolute magnetic field value in an active region.
A widely used alternative W-NLFFF code has been 
created by Wiegelmann et al.~(2006, 2012), which we used for comparison
in Paper I also.  For the subset of all (11) X-class flares, a mean
ratio of $E_p^{phot}/E_p^{cor} = 1.05 \pm 0.33$ was found.
Here we rename the photospheric NLFFF code of Wiegelmann
by $E_W=E_p^{phot}$, and the coronal (vertical current approximation)
code by $E_{VCA3}=E_p^{cor}$, which yields an inverse ratio of 
$E_{VCA}/E_{W} = 0.95 \pm 0.31$ for the potential field energy. 
Using the improved new VCA3-NLFFF code, we find a mean ratio of
\begin{equation}
	{ E_{VCA3} \over E_W } = 0.99 \pm 0.21 \ ,
\end{equation}
for the 11 cases shown in Fig.~(4-5), measured at the peak time
of the free energy (which generally occurs between the flare start
and flare peak in soft X-rays (as defined by GOES). Our new result
is consistent with the old results of $E_{VCA}/E_W = 0.95 \pm 0.31$,
and yields an overall mean value that is closer to unity (as expected
for a perfect absolute calibration of the code), and moreover has
a smaller scatter ($\sigma=\pm0.21$) than the old code ($\sigma=\pm0.31$). 
Therefore, the new VCA3-NLFFF code appears to retrieve the absolute
magnetic field more accurately than the previous VCA-NLFFF code, 
thanks to the inclusion of the poloidal field component $(B_\theta)$
and the empirical calibration factor $q_B=0.8$ (Eq.~32). 

However, it is not entirely clear what affects the accuracy in the measurement
of the magnetic field strength. Since we compare the potential field
strength only, which excludes the nonlinear effects of any NLFFF code,
we suspect that the accuracy is affected by the decomposition of the 
magnetograms into unipolar magnetic charges. For axi-symmetric magnetic
field structures, such as they occur above a sunspot with spherical
symmetry, the model of buried unipolar charges is expected to be most
accurate. If two sunspots with opposite magnetic polarity appear to be 
spatially separated on the solar surface, the magnetic flux should 
be perfectly conserved. However, if two sunspots with opposite magnetic
polarity overlap each other, part of the oppositely-directed magnetic
polarity cancel out, so that the signed flux is not conserved.
This interpretation predicts that the unsigned magnetic flux should be
conserved for separated sunspots, while the unsigned flux is likely
to be under-estimated for spatially overlapping sunspot pairs.
We investigated whether systematic errors in the magnetic field
reconstruction depend on the separation of the main sunspots,
or on the heliographic longitude, but did not found a systematic
pattern. Nevertheless, the empirical correction factor
$q_e^{model}=0.8^2=0.64$ (Eq.~32) for magnetic energies,
which corresponds to a correction factor of 
$q_B^{model}=0.8$ for magnetic fields (Eq.~32), 
yields a statistically accurate correction that agrees with other
NLFFF codes within a few percents (Eq.~37). 

\subsection{	Magnetic Energy Ratios				}

A key result of this study is the more accurate measurement of
the four types of magnetic energies, i.e. the potential energy
$E_p(t)$, the nonpotential energy $E_{np}(t)$, the free energy $E_{free}(t)$,
and the total dissipated magnetic energy $E_{diss}$ during a flare.
As representative values we use the time averages of the potential energy,
\begin{equation}
	E_p=\sum E_p(t)/n_t \ ,
\end{equation}
and the nonpotential energy,
\begin{equation}
	E_{np}=\sum E_{np}(t)/n_t \ .
\end{equation}
The free energy, which is generally defined as
\begin{equation}
	E_{free}(t)=E_{np}(t)-E_p(t) \ ,
\end{equation}
contains two time-dependent components: one is the time intervals with
positive increases of the free energy, and one is the time intervals with
negative decreases of the free energy. We interpret the positive increases
as injection of free energy from the photosphere or chromosphere 
(or new emerging current-carrying flux), while
the negative decreases represent the dissipation of magnetic energy in
the solar corona during a flare magnetic reconnection process. 
Thus we can define the time evolution of magnetic energy dissipation
by a cumulative function that contains only time intervals with negative
decreases (similar to method in Paper I),
\begin{equation}
	E_{cum}(t_i) = E_{cum}(t_{i-1}) - 
		\left( [E_{free}(t_i)-E_{free}(t_{i-1})] > 0 \right)
	\ ,
\end{equation} 
This characterization in terms of a cumulative time evolution function,
which is monotonically decreasing with time, allows us to separate the
internal energy dissipation from the external energy injection of
additional energy. The time intervals of energy dissipation are
rendered with hashed areas in the right-hand panels of Figs.~(6) and (7).
As representative values of the free energy we use the initial maximum
of the cumulative time evolution function,
\begin{equation}
	E_{free} =max[E_{cum}(t)]) \ ,
\end{equation}
and as a representative value of the dissipated energy the difference
between the initial maximum and the final minimum of the cumulative
time evolution function,
\begin{equation}
	 E_{diss}=max[E_{cum}(t)] - min[E_{cum}(t)] \ .
\end{equation}
This definition limits the dissipated energy to 
$E_{diss} \le E_{free}$ at the upper end and to $E_{diss} > 0$ at the
lower end, and therefore avoids unphysical solutions with 
$E_{diss} > E_{free}$.

In Fig.~8 we show the various energy correlations as a function of the
potential energy (Fig.~8a,b,c) or the free energy (Fig.~8d), where we
obtain the following linear regression fits in log-log space, 
\begin{equation}
	\left( {E_{np} \over 10^{30}\ {\rm erg}} \right) = 0.96 
	\left( {E_{p} \over 10^{30}\ {\rm erg}} \right)^{1.019} \ ,
\end{equation}
\begin{equation}
	\left( {E_{free} \over 10^{30}\ {\rm erg}} \right) = 0.018 
	\left( {E_{p} \over 10^{30}\ {\rm erg}} \right)^{1.261} \ ,
\end{equation}
\begin{equation}
	\left( {E_{diss} \over 10^{30}\ {\rm erg}} \right) = 0.0065 
	\left( {E_{p} \over 10^{30}\ {\rm erg}} \right)^{1.320} \ ,
\end{equation}
\begin{equation}
	\left( {E_{diss} \over 10^{30}\ {\rm erg}} \right) = 0.43 
	\left( {E_{free} \over 10^{30}\ {\rm erg}} \right)^{1.044} \ .
\end{equation}
These nonlinear relationships can be approximated by 
averaged energy ratios, for which we find the following means and 
standard deviations of (Fig.~8, dotted lines),
\begin{equation}
	\left( {E_{np} \over E_p} \right) \approx 1.10 \pm 0.03 \ ,
\end{equation}
\begin{equation}
	\left( {E_{free} \over E_p} \right) \approx 0.12 \pm 0.03 \ ,
\end{equation}
\begin{equation}
	\left( {E_{diss} \over E_p} \right) \approx 0.07 \pm 0.03 \ ,
\end{equation}
\begin{equation}
	\left( {E_{diss} \over E_{free}} \right) \approx 0.60 \pm 0.26 \ .
\end{equation}
Note that the uncertainties of these energy ratios are significantly 
smaller than in the
previous study (Paper I), which indicates less scatter due to systematic
errors, especially for those with unphysical solutions $E_{diss} > E_{free}$
in the previous study. As a rule of thumb we obtain the main results that
the nonpotential energy is roughly 110\% of the potential energy,
the free energies and dissipated energies amount typically to $\approx 10\%$ 
of the potential energy, and the dissipated energy is about half the free energy.
These results are consistent with those in the previous study 
(Fig.~13 in Paper 1), but in addition display tighter correlations, 
smaller spreads in the energy ratios, and elimination of unphysical 
solutions.

\subsection{	Energy Closure in Flares		}

One of the most important tasks in our study is the energy partition
and energy closure in flares. Quantitative information on different
forms of energies and their partition in flares and CMEs became 
more available lately (Emslie et al.~2012; Warmuth and Mann 2016a, 2016b;
Aschwanden et al.~2014a, 2015a, 2016a; 2017; 2019; Aschwanden 2016a, 2017;
Aschwanden and Gopalswamy 2019). Virtually no statistical study
on flare energies existed 5 years ago (Aschwanden 2004, 2019a). 
In Fig.~9 we show scatterplots of the
various forms of energies as a function of the dissipated magnetic
energy, which supposedly is the upper limit of any energy that can
be dissipated in a flare. The panel in Fig.~9a shows a comparison
between the old (VCA-NLFFF) and the new (VCA3-NLFFF) code, being
a factor of 3 lower with the new code. Nevertheless, the 
averaged ratios of the other forms energies are all below unity,
see Eqs.~(48-51), which means that solar flares provide sufficient
free (magnetic) energy to supply the nonthermal energy in accelerated
electrons and ions, as well as direct heating and expulsion of coronal
mass ejections (CMEs) (Eqs.~48-51), yielding the mean ratios,
\begin{equation}
	\left( {E_{nt,e} \over E_{diss}} \right) \approx 0.34 \pm 0.09 \ ,
\end{equation}
\begin{equation}
	\left( {E_{nt,i} \over E_{diss}} \right) \approx 0.11 \pm 0.03 \ ,
\end{equation}
\begin{equation}
	\left( {E_{dir} \over E_{diss}} \right) \approx 0.36 \pm 0.09 \ ,
\end{equation}
\begin{equation}
	\left( {E_{cme} \over E_{diss}} \right) \approx 0.31 \pm 0.08 \ .
\end{equation}
We can (now test the energy closure by summing the primary energy release
processes in $E_{tot}$, where the thermal energy $E_{th}$ is defined as
the sum of nonthermal heating (by electron and ion precipitation) and
direct heating,
\begin{equation}
	E_{tot} =  E_{nt,e} + E_{nt,i} + E_{dir} + E_{cme} = E_{th} + E_{cme} \ ,
\end{equation}
for which we obtain a ratio of (Fig.~9f), 
\begin{equation}
	\left( {E_{tot} \over E_{diss}} \right) = 0.79 \pm 0.12 \ .
\end{equation}
A pie chart of this energy partition is shown in Fig.~10c, which
significantly differs from our previous study (Fig.~10b) or the
study of Emslie et al.~(2012) (Fig.~10a). In Fig.~11 we depict the relative
energy partitions of 12 individual flares, which have an energy
closure within a factor of $1/4 \lapprox E_{tot}/E_{diss} \lapprox 4.0$. 
These examples document large differences between individual
flares that need to be explored in more details in future studies.

\subsection{	Energy Size Distributions		}

In Fig.~12 we show the size distributions of the same 6 forms of
energies as in the scatterplots of Fig.~9. Ignoring the flattening
turnover at the lower side of the logarithmic size distributions,
which results from undersampling of weak flares with class M$<$1.0,
all six size distributions exhibit a power law distribution 
$N(E) \propto E^{-\alpha}$, with a slope of $a \approx 1.6-1.8$,
which agrees well with predictions of self-organized criticality
(SOC) models. A statistical fractal-diffusive avalanche model
of a slowly-driven SOC system with Euclidean dimension $S=3$
and fractal dimension $D_S=(1+S)/2=2$ predicts a power law slope of
(Aschwanden 2012),
\begin{equation}
	\alpha_P = 2 - {1 \over S} = {5 \over 3} \approx 1.67 \ ,
\end{equation}
for the peak energy dissipation rate $P$, and
\begin{equation}	
	\alpha_E = 1 + {(S-1) \over (D_S + 2)} = {3 \over 2} = 1.5 \ ,
\end{equation}
for the dissipated energy distribution. The accuracy of the power law
slope is on the order of $\sigma_\alpha \approx 0.1-0.2$, limited by the
relatively small number of analyzed events ($N \approx 10^2$), which
yields inertial ranges of 1-2 decades in the size of energies only. 
The values, however, are close to power law slopes in larger samples,
e.g., $\alpha_E = 1.53 \pm 0.02$ for nonthermal (electron) energies
in flares (Crosby et al.~1993). 

\section{	DISCUSSION					}

Here we discuss some consequences of our magnetic energy measurements,
which have a bearing on the capability of coronal energy storage
(Section 5.1), the time scale separation of energy storage and dissipation
(Section 5.2), the conservation of magnetic energy 
(Section 5.3), the critical twist and free energy before flaring
(Section 5.4), and forecasting of the flare magnitude (Section 5.5).

\subsection{	Coronal Energy Storage 				}

Since we have more accurate measurements of the time evolution of
free energy, before, during, and after flares, we briefly revisit 
the old issue of energy storage in the solar corona. 
Rosner and Vaiana (1978) envisioned a unified model of cosmic flare
transients, where the time evolution of energy storage and release
can be derived from the event occurrence frequency distribution. Simply put,
energy is stored in the solar corona with an exponential growth rate,
which produces a power law distribution of flare energies, if the
time intervals between two subsequent flares are governed by a
stochastic (random) waiting time distribution. However, the postulated
correlation between the waiting time and the intermittently released
flare energies has never been confirmed by observational statistics
in solar flares (Crosby et al.~1998; Wheatland 2000; Lippiello et al.~2010), 
and time scale arguments were brought forward that indicate that such 
a storage model is inconsistent with solar flare energy storage in 
coronal magnetic fields (Lu 1995).  

If the coronal energy storage model would be true, we would expect a high
level of stored free energy shortly before the flare, which then
reduces to a lower level after the flare, manifesting an approximate
step-function in the free energy $E_{free}(t)$. Although we do find
an unambiguous step function in the free energy in some cases, 
with equal magnitude measured by both the W-NLFFF and VCA3-NLFFF code
(e.g., flare \#67 in Fig.~6d), the step-function does often not has the
same magnitude in both codes (e.g., flare \#384 in Fig.~7e or
flare \#220 in Fig.~7a), or the step function is accompanied by
other impulsive magnetic energy increases after
the flare peak time (e.g., flare \#66 in Fig.~6c, or flare \#344
in Fig.~7b). These erratic fluctuations in the free energy $E_{free}(t)$
indicate that a simple energy storage model with a monotonic increase
before the flare peak, and with an irreversible decrease after the 
flare peak time (defined by the maximum of the GOES time derivative),
is not always consistent with the observations, although we detect
some level of energy decrease in all cases (Figs.~6 and 7).
Consequently we should generalize the step function scenario to a
more comprehensive model that allows coronal energy storage and
dissipation on time scales that are even shorter than the flare
duration. 

We illustrate the impulsiveness of free energy storage
and dissipation in Fig.~13 for the case of the longest observed
flare in our sample, with a total duration of $\tau_{flare}=4.1$ hrs. 
In this case we see numerous decreases of the free energy (13 red hatched
areas in Fig.~13 middle panel). Clearly the storage and dissipation
of free energy varies on much shorter time scales than the flare
duration, in contrast to the coronal storage model of Rosner and
Vaiana (1978), where energy storage times are required to be
much longer than the flare duration, continuously accumulating 
during two subsequent large flares (which may not repeat until
time scales of days to weeks according to observations). 

\subsection{	Time Scale Separation of Energy Storage and Dissipation  }

The previous discussion raises the issue of time scale
separation (see also Section 2.13 in Aschwanden et al.~2016b). 
Self-organized criticality models assume a time scale
separation between the energy build-up or storage time scale 
$\tau_{storage}$, and the duration $\tau_{diss}$, i.e.,
$\tau_{storage} \gg \tau_{diss}$, in order to enable a slowly-driven 
system that produces a power law distribution of dissipated energies,
as originally proposed for sandpile avalanches by Bak et al.~(1987).
The waiting times between subsequent avalanches are assumed
to be a stochastic (random) process and to produce a Poisson distribution
(Wheatland et al.~1998).
The same criterion of time scale separation has been applied to the energy
storage model of Rosner and Vaiana (1978), i.e., $\tau_{storage} 
\gg \tau_{diss}$, but the size distribution of the avalanches is 
assumed to follow some scaling with the waiting times, which would
constitute a deterministic system, where the flare magnitude could
be predicted by the waiting time, which however is not the case 
(see the following discussion in Section 5.5). Moreover, the time
scales of energy build-up or storage, as well as the time scales
of energy dissipation, are observed to be much shorter (in the order
of minutes) than the flare duration (up to 4 hrs for the case shown
in Fig.~13), and thus the observed time scale ranges are opposite
to the model of Rosner and Vaiana (1978), as well as opposite to
slowly-driven SOC models, which leaves us with ``fast-driven'' or
``strongly-driven'' SOC systems (Aschwanden et al.~2016b). 

\subsection{	Energy Conservation During Solar Flares		}

Earlier estimates of flare energies assumed that the free 
(magnetic) energy has a high level before the flare, and then drops
like a step-function to a lower level after the flare, which
implies energy conservation of the free energy in the solar
corona, i.e., 
\begin{equation}
	E_{free}(t_{before}) - E_{diss} = E_{free}(t=t_{after}) \ .
\end{equation}
Since we developed tools to calculate the nonpotential energy
and the free energy $E_{free}(t)$ (e.g., with the W-NLFFF or 
VCA3-NLFFF code), we can apply the principle of energy conservation
in order to determine the combined energy dissipation of
the primary processes operating in flares (Eq.~56), which we
estimated from the cumulative energy decrease function
$E_{cum}(t)$ (by extracting it from the free energy $E_{free}(t)$
as shown in Figs.~6-7). However, since the energy dissipation
often does not follow a single step-function, but rather multiple
step-functions (up to $\approx 14$ episodes in the long-duration
flare shown in Fig.~13), we should introduce a more generalized
energy conservation formalism that includes multiple 
step functions, 
\begin{equation}
	\sum E_{free,pos}(t) \Delta t - E_{diss} 
	= \sum E_{free,neg}(t) \Delta t \ ,
\end{equation}
where $E_{free,pos}(t)$ and $E_{free,neg}(t)$ encompass time intervals
with positive 
and negative changes of the free energy. In this study we measured
the cumulative negative changes of the free energy, 
$E_{cum}(t)=E_{free,neg}(t)$,
but did not measure the sum of the positive free energy changes,
since there is some ambiguity about the flare duration and the
definition of the flare-associated pre-flare and post-flare phases.

Nevertheless, our analysis indicates that the flare-associated
changes of the free (magnetic) energy is often more complex than a 
single step-function. Moreover, since there is no simple monotonic
increase of stored energy in the corona, but rather intermittent
pulses of energy injection with rapid dissipation during flare
phases, probably of chromospheric origin, energy conservation
cannot simply be deduced from the coronal (non-potential) magnetic
field alone, but has to include also energy input from the chromosphere.
In this sense magnetic energy is not conserved for coronal
contributions alone, but requires the knowledge of the 
energy contributions from the chromosphere also.  

\subsection{	Critical Twist and Free Energy 		}

A perhaps surprising result is the small ratio of the free 
magnetic energy to the potential energy, i.e., 
$E_{free}/E_p=0.12\pm0.03$ (Fig.~8b; Eq.~49).
Why is the available free energy an order of
magnitude smaller than the total potential field energy ?
According to the helically twisted geometry of a flux tube,
this energy ratio can be translated into a geometric angle
between the potential and nonpotential field components,
which is also known as helical (or azimuthal) twist angle,
\begin{equation}
	{E_{free} \over E_p} 
	= {B_\varphi^2 \over B_p^2} = (\tan \mu)^2 \ .
\end{equation}
According to our measurements with $E_{free}/E_p \approx 0.12\pm0.03$,
the energy ratio corresponds to a helical twist angle of
\begin{equation}
	\mu = \arctan \left[ \left( {E_{free} \over E_p} \right)^{1/2} \right] \ ,
\end{equation}
yielding a value of $\mu = 19^\circ \pm 2^\circ$. Thus we suspect 
that helical twisting is occurring until a critical angle of 
$\mu_{crit} \approx 19^\circ$ is reached, while further accumulation
of free energy by helical twisting is inhibited by some stabilizing
force, or by the onset of an instability. 

The number of twisting turns in a flux tube can be expressed by
\begin{equation}
	\tan \mu = {N_{twist} \ 2 \pi r \over L} 
		 \approx 6.3\ N_{twist} \left({r \over L}\right) \ .
\end{equation}
where $r$ is the flux tube radius and $L$ is the loop length. 
For instance, a critical twist angle of $\mu=19^\circ$ and 
an aspect ratio of $(r/L)=(1/20)$ yields a number of one
twist turn, i.e., $N_{twist} \approx 1.0$, which is the
approximate threshold for the kink instability
(Priest 2014; Hood and Priest 1979; T\"or\"ok et al.~2003; 
Kliem and T\"or\"ok 2006; Kliem et al.~2014).
Hence, a constant value of the critical helical twist angle
can explain the proportionality between the free energy
$E_{free}$ and the potential energy $E_p$ found here (Eq.~49),
as well as the prediction that the free energy scales
with the helical twist angle $\mu$ (Eq.~63).  

\subsection{		Flare Magnitude Prediction			}

The performance of solar flare forecasting methods has been
diligently tested, but there is no consensus about the best method
(Schrijver 2007; Georgoulis and Rust 2007; Leka and Barnes 2003;
Barnes et al.~2007, 2016; Barnes and Leka 2008; Bloomfield et al.~2012;
Colak and Qahwaji 2009; Li et al.~2007; Ahmed et al.~2013;
Falconer et al.~2003, 2011, 2012; Mason and Hoeksema 2010; Cui et al.~2006;
Bobra and Ilonidis 2016; Nishizuka et al.~2017; Jonas et al.~2018;
B\'elanger et al.~2007; Strugarek and Charbonneau 2014).
A question of general interest is whether the evolution of
the free energy, $E_{free}(t)$ can be used for flare forecasting?
The free energy has been found to be one of the best correlated
flare indicators (Bobra and Couvidat 2015). However, the free energy
calculated in Bobra and Couvidat (2015) is derived from photospheric
magnetograms, which measure the longitudinal (line-of-sight) component
$B_z$ and the transverse components $B_x$ and $B_y$ in the
non-forcefree photosphere (Metcalf et al.~1995).  Another well-correlated
parameter that can be calculated with our VCA3-NLFFF code includes
the total unsigned vertical current. 

Our statistics with enhanced accuracy provides some useful information
for forecasting of the magnitude of solar flares. The tight correlations
between the four magnetic energy parameters found here (i.e., the potential,
non-potential, free energy, and flare-dissipated energies) (Fig.~8)
enables us to predict the flare magnitude from the knowledge of the
potential field of an active region. The potential field is easy to
calculate and is slowly-varying in an active region, often being
stable over several days (Sun et al.~2012). 
From another statistical study it was concluded that 
currents associated with coronal nonpotentiality have a characteristic
growth and decay time scale of $\approx 10-30$ hrs (Schrijver
et al.~2005; Schrijver 2016).
The free energy is confined to a 
relative small fraction of $12\%\pm 3\%$ of the potential energy
(Eq.~49). Also the flare-dissipated energy is found in a relatively
small range of $7\%\pm 3\%$ (Eq.~50), from which we can predict an upper limit
of the dissipated energy a few days ahead of the flare. 
We can not predict the specific onset 
time of the flare, but we can provide an upper limit of the largest 
flare occurring the next few days. The uncertainty of our 
prediction method, based on the statistical ratio found here for 
M- and X-class flare events, i.e., $E_{diss}/E_p=7\%\pm3\%$ (Eq.~50), 
amounts to a factor of $10\%/7\%=1.4$. The largest
nonpotential energy we measured in this study is
$E_{np} \approx 6 \times 10^{33}$, from which we predict
a maximum dissipated flare energy of $7\%\pm3\%$, or 
$E_{diss}=4.2 \pm 1.8 \times 10^{32}$ erg, similar to 
our actual measurement of $E_{diss}^{obs}=6.4 \times 10^{32}$ erg
(Table 2). 

\section{		CONCLUSIONS				}

In this study and in Paper I we follow the approach of modeling the
magnetic energies in flaring active regions by calculating approximative
analytical solutions of non-linear force-free fields in terms of
the vertical-current approximation, using the observational data 
of projected coordinates $[x(s), y(s)]$ of (automatically traced) 
coronal loops from EUV (AIA/SDO) images, and the line-of-sight 
component $B_z(x,y)$ from photospheric (HMI/SDO) magnetograms. 
In Paper I we derived an analytical solution of the magnetic field 
components $[B_r({\bf x}), B_{\varphi}({\bf x})]$ in spherical coordinates 
$[r, \varphi, \theta]$, neglecting the poloidal field component 
($B_\theta({\bf x})=0$). In this Paper IX, we refine the analytical solution 
by including a solution of the poloidal field component 
($B_{\theta}({\bf x}) \ne 0$), which yields more accurate values of the
magnetic field ${\bf B}({\bf x, t})$ and (non-potential) magnetic 
energies $E_{np}({\bf x, t})$. The new analytical solution, which
is calculated with the VCA3-NLFFF code, is of second-order accuracy
in the divergence-freeness condition, and of third-order accuracy in the
force-freeness (or solenoidal) condition. The main new results and 
conclusions are:

\begin{enumerate} 
\item{{\bf The Gold-Hoyle flux rope test:} Highly twisted flux ropes
with a large number of helical windings are thought to drive the
expulsion of coronal mass ejections from a coronal flare site out
to the heliosphere (Gold and Hoyle 1960). Such a highly twisted
structure with 5 windings could be reproduced with the old VCA-NLFFF
code (Aschwanden 2013a), while the new VCA3-NLFFF code could only 
produce magnetic field lines with less than a half winding turn
(Aschwanden 2019), but this inability is consistent with the expected
non-existence of nonlinear force-free field solutions 
at $N_{twist} \gapprox 1$ due to the kink instability criterion. 
However, such
Gold-Hoyle structures may exist as transients in interplanetary space,
where force-freeness is not required.}

\item{{\bf Accuracy of the VCA3-NLFFF code:} The accuracy of a
magnetic field measurement ${\bf B}({\bf x})$ with the VCA3-NLFFF code
is limited in zones where magnetic concentrations with opposite magnetic
polarity overlap, so that the signed magnetic flux partially cancels out.
We find that this results into an underestimate of the magnetic field 
by a statistical factor of $q_B \approx 0.8$. Correcting for this
statistical factor yields an agreement of the magnetic (nonpotential) energy
within a mean and standard deviation of $E_{VCA3}/E_W=0.99\pm0.21$
with respect to the Wiegelmann (W-NLFFF) code.} 

\item{{\bf Time evolution of the magnetic field:} In the simplest concept
we would expect that the free energy has a steady, relatively high value 
before a flare, then decreases like a step-function, and reaches a relatively 
low value after the flare. While this single step-function scenario 
approximates the measured free energy $E_{free}(t)$ quite well 
in some cases, the detailed evolution of the free energy exhibits multiple
short-term increases, followed by decreases on time scales that
are shorter than the flare duration, up to 13 such episodes during
a ($\approx 4$ hour) long-duration flare. The energy evolution
during flares thus exhibits an intermittent and impulsive behavior
of energy build-up and dissipation that is more complex than a single 
step-function.}

\item{{\bf Magnetic energy ratios:} Defining a cumulative (monotonically
decreasing) function of negative energy steps during the flare duration, 
we obtain a measure for the 
evolution of the free energy $E_{free}(t)$ during a flare, bound by the  
energy difference between the flare start and end. We find that the
dissipated energy amounts to a statistical ratio of $E_{diss}/E_{free}
=0.60\pm0.26$ with respect to the free energy, or $E_{diss}/E_{p}=0.07\pm0.03$ 
with respect to the potential energy of a flaring active region. The latter
ratio is useful for predicting upper limits of flare energies.}

\item{{\bf Energy closure:} The total dissipated energy in a flare
($E_{tot}$) is converted into nonthermal energy of accelerated
electrons $(E_{nt,e})$ and ions ($E_{nt,i})$, direct heating of coronal plasma
$(E_{dir})$, and possible launch of a CME $(E_{cme})$, which can be
expressed also by the sum of the thermal energy $E_{th}$ (in a secondary energy 
conversion step) and the kinetic CME energy $E_{cme}$. We find an 
energy closure of $E_{tot}/E_{diss}=0.79\pm0.12$, which compares favorably
with earlier work, i.e., $E_{tot}/E_{diss}=0.87\pm0.18$ (Aschwanden
et al.~2017; Paper V). The energy closure in individual events, however, varies 
by a much larger factor, which indicates large uncertainties in individual 
energy measurements that need to be investigated in further detail.}

\item{{\bf Coronal energy storage:} The intermittent and impulsive
behavior of the time evolution of the free energy $E_{free}(t)$
contradicts the single-step energy dissipation scenario of 
Rosner and Vaiana (1978). We find up to 14 episodes of
consecutive free energy increases and decreases in a (4-hour long-duration)
flare event, with a cadence (of 6 min) that is only a factor 3 shorter
than the mean period of the free energy fluctuations. Long-duration energy 
storage with a singular energy dissipation phase thus cannot explain 
the observed time variability of the free energy.}
  
\item{{\bf Time scale separation:} The increase of free energy in a
flaring region, which is expected over a relatively long duration
(compared with the flare duration), is found to fluctuate almost
erratically before, during, and after flares. These short-term
fluctuations thus violate the time scale separation $\tau_{flare}
\ll \tau_{storage}$, which is also expected in slow-driven
self-organized criticality (SOC) models, and instead require a
``fast-driven'' or ``strongly-driven'' model (Aschwanden et al.~2016b),
which might produce significant deviations from power law size
distribution. The size distributions observed here, however, 
preserve the power law distribution and match the predicted
slopes of $\alpha \approx 1.5-1.7$ (Fig.~12).}

\item{{\bf Energy conservation:} The simple step-function scenario
of coronal energy build-up and dissipation implies energy conservation
of the coronal magnetic field, where the free energy before a flare
matches the free energy after the flare plus the amount of dissipated
energy associated with the primary energy conversion processes
(Eq.~56). The mismatch of the single step-function scenario, however,
suggests that energy is externally injected into the coronal flare
region (such as vertical currents from the chromosphere), in an 
intermittent way, since the storage process seems to be independent 
of the dissipation process. Consequently, energy conservation during 
flares requires the knowledge of external energy injections also.}

\item{{\bf Critical twist and kink instability:} Fitting the
vertical-current approximation NLFFF model to coronal loops
yields the ratio of the helically twisted azimuthal magnetic
field component $B_\varphi ({\bf x})$ to the potential radial 
field component $B_r ({\bf x})$, which implies a helical twist
angle of $\mu=19^\circ\pm2^\circ$ in our data, or a winding number 
of $N_{twist} \lapprox 0.5$ (Aschwanden 2019b). This relatively low
value is also compatible with the critical twist angle or threshold
of the kink instability at $N_{twist} \gapprox 1.0$ 
(T\"oroek et al.~2003). This instability 
limit sets also an upper limit for the ratio of the free energy to 
the potential free energy, and this way constrains a useful 
criterion for the prediction of flare magnitudes.}

\item{{\bf Flare magnitude prediction:} We found tight correlations
between the (potential, non-potential, free, and dissipated)
magnetic energy parameters. The mean ratio between the dissipated
flare energy $E_{diss}$ and the potential energy $E_{free}$,
for instance, has a relatively narrow ratio of
$E_{diss}/E_p=0.07\pm0.03$, which implies a standard deviation
by a factor of 1.4 only. Since the potential energy of an
active region is generally slowly-varying over the course of a
few days, $E_{p}(t)$ is almost constant (Sun et al.~2012), and upper limits 
$E_{diss} \lapprox (0.07 \pm 0.03) \times E_p$ can be predicted 
for any flare a few days ahead.}

\end{enumerate}
We carried out magnetic modeling with both the VCA-NLFFF and the
W-NLFFF code, in both Paper I and in this present study (Paper IX).
The comparisons help us to better understand the absolute and relative 
uncertainties. It is gratifying to see that we obtain relatively 
consistent results, regardless whether we use transverse magnetic
field components $(B_x, B_y)$ (W-NLFFF) or loop coordinates
$[x(s), y(s)]$ (VCA-NLFFF). Theoretically, the VCA-NLFFF code
is preferable because it circumvents the non-forcefreeness
issue in the photosphere (Metcalf et al.~1995; DeRosa et al.~2009), 
and instead uses geometric constraints
of coronal loops, while the W-NLFFF code executes coronal field
extrapolations rooted in photospheric non-forcefree fields. 
Future magnetic field computation methods may combine the advantages
of both types of codes. 

\acknowledgements
The author thanks for helpful discussions with Marc DeRosa,
John Serafin, and Wei Liu.
Part of the work was supported by
NASA contract NNG 04EA00C of the SDO/AIA instrument and
the NASA STEREO mission under NRL contract N00173-02-C-2035.

\section*{APPENDIX A}

In this Appendix A we derive our new analytical solution 
of a divergence-free and force-free magnetic field (used in the
new VCA3-NLFFF code), which represents a more accurate solution than 
the original approximation derived in Aschwanden (2013a), where
the poloidal magnetic field component was neglected, i.e., $B_\theta = 0$. 
Thus we start from the original approximation (Eqs.~A1, A2, A3, A4),
$$
	B_r(r, \theta) = B_0 \left({d^2 \over r^2}\right)
	{1 \over (1 + b^2 r^2 \sin^2{\theta})} \ , 
	\eqno(A1)
$$
$$
	B_\varphi(r, \theta) = 
	B_0 \left({d^2 \over r^2}\right)
	{b r \sin{\theta} \over (1 + b^2 r^2 \sin^2{\theta})} \ ,
	\eqno(A2)
$$
$$
	B_\theta(r, \theta) \approx 0 \ ,
	\eqno(A3)
$$
$$
	\alpha(r, \theta) \approx {2 b \cos{\theta} \over 
	(1 + b^2 r^2 \sin^2{\theta})}  \ .
	\eqno(A4)
$$
{\bf where the constant $b$ is defined in terms of the number
of full twisting turns $N_{twist}$ over the loop length $L$,   
i.e., $b = 2 \pi N_{twist}/L$.} 
From the second equation of the forcefree condition (Eq.~16) we can
express the poloidal field component explicitly,
$$
	B_\theta (r, \theta ) = - {1 \over r \alpha} 
		{\partial \over \partial r} ( r B_\varphi ) \ ,
	\eqno(A5)
$$
where we can insert the azimuthal field component $B_\varphi (r, \theta)$
(Eq.~A2), the nonlinear $\alpha$-parameter, $\alpha(r, \theta)$ (Eq.~A4),
calculate the radial derivative $\partial (r B_\varphi ) / \partial r$, 
and obtain this way an explicit expression for the
poloidal field component,
$$ 
	B_\theta(r, \theta) = 
	B_0 \left({d^2 \over r^2}\right)
	{b^2 r^2 \sin^3(\theta) \over
	(1 + b^2 r^2 \sin^2 \theta )} {1 \over \cos \theta}
	\ .
	\eqno(A6)
$$
which represents a more accurate approximation than the original
approximation with $B_\theta=0$. For numerical calculations we find
that the expression (A6) is numerically unstable, but can be fully 
stabilized by approximating the cosine term with $\cos(\theta)=1$,
without losing significant accuracy.

A second task is the accuracy of the divergence-free condition (Eq.~10),
which for axi-symmetric fields 
$(\partial / \partial \varphi = 0)$ reduces to Eq.~(14), 
$$
        {1 \over r^2} {\partial \over \partial r} (r^2 B_r)
        + {1 \over r \sin{\theta}} {\partial \over \partial \theta}
        (B_\theta \sin{\theta}) = 0
        \ .
	\eqno(A7) 
$$
We can insert the expressions for the radial magnetic field component
$B_r(r, \theta)$ (Eq.~A1), the poloidal magnetic field component
$B_\theta(r, \theta)$ (Eq.~A6), and calculate the derivatives
$\partial (r^2 B_r) / \partial r$ and  
$\partial (B_\theta \sin \theta) / \partial \theta $, which after
some lengthy algebra leads to the expression,   
$$
	\nabla \cdot {\bf B} = 
	{B_0 d^2 \over r^3}
	\left[ {2 b^2 r^2 \sin^2\theta \cos^2 \theta
	      +b^2 r^2 \sin^4\theta
		-2 b^4 r^4 \sin^4 \theta \cos^2 \theta
		+ b^4 r^4 \sin^6 \theta \over
		\cos^2 \theta ( 1 + b^2 r^2 \sin^2 \theta) ^2 } \right] 
	\approx \left( {B_0 d^2 \over r^3} \right) \ O(b r \sin \theta)^2
	\ ,
	\eqno(A8)
$$
which is of second-order in the argument $(b r \sin \theta)$,
similar to the original approximation (Aschwanden 2013a). This means that
the approximation is most accurate for a weakly nonlinear forcefree
magnetic fields with $(b r \sin \theta) \ll 1$ or $\alpha \ll 1$.  
Note that the divergence of the potential field can be obtained from Eq.~(A8)
by setting $b=0$, 
$$
	\nabla \cdot {\bf B} = 
	\left({B_0 d^2 \over r^3}\right) \times O = 0 
	\ . 
	\eqno(A9) 
$$

A next task is the assessment of the accuracy of the force-free
condition (Eq.~9 or 11-13). For the radial component 
$[\nabla \times {\bf B}]_r$ (Eq.~11), and applying axi-symmetry
$(\partial / \partial \varphi=0)$, Eq.~11 reduces to,
$$
	\left[ \nabla \times {\bf B} \right]_r =
 	{1 \over r \sin{\theta}}
 	\left[{\partial \over \partial \theta}
 	(B_\varphi \sin{\theta} \right] 
	= \alpha B_r \ .
	\eqno(A10)
$$
Inserting the azimuthal field component $B_\varphi(r, \theta)$ (A2) 
and calculating the derivative $\partial \theta
(B_\varphi \sin \theta) / \partial \theta$, 
we find that all terms in this expression
cancel out, so that this forcefree condition is exactly fulfilled,
$$
	\left[ \nabla \times {\bf B} \right]_r = 0 \ ,
	\eqno(A11)
$$
which was also the case in the old VCA-NLFFF code, as given in Aschwanden (2013a). 
The second vector component of the forcefree condition, i.e., 
$[\nabla \times {\bf B}]_\theta$ (Eq.~12), with axi-symmetry applied
$(\partial / \partial \varphi = 0)$, simplifies to
$$
	\left[ \nabla \times {\bf B} \right]_\theta =
	- \left[ {1 \over r} {\partial \over \partial r} (r B_\varphi) \right]
	= \alpha B_\theta \ .
	\eqno(A12)
$$
Inserting the magnetic field components 
$B_\varphi (r, \theta)$ (Eq.~A2), $B_\theta (r, \theta)$ (Eq.~A6), 
the nonlinear parameter $\alpha (r, \theta)$ (Eq.~A4), and calculating
the derivative $\partial (r B_\varphi) / \partial r $ yields then the
expression,
$$
	\left[ \nabla \times {\bf B} \right]_\theta =
	\left( B_0 {d^2 \over r^3} \right)
	{(-2 b^5 r^5 \sin^5 \theta) \over
	(1 + b^2 r^2 \sin^2 \theta)^2 }
	\approx 
	\left( {B_0 d^2 \over r^3} \right)
	O(b r \sin \theta)^5 
	\ ,
	\eqno(A13)
$$
which is of fifth-order accuracy in the argument $(b r \sin \theta)$.

Finally we assess the accuracy of the third vector component of the
forcefree condition, $[\nabla \times {\bf B}]_r$ (Eq.~11).
The axi-symmetry $\partial / \partial \varphi = 0$ simplifies the
expression Eq~(11) to,
$$
	\left[ \nabla \times {\bf B} \right]_r =
 	{1 \over r \sin{\theta}}
 	\left[{\partial \over \partial \theta}
 	(B_\varphi \sin{\theta}) \right] =  \alpha B_r \ ,
	\eqno(A14)
$$ 
Inserting the magnetic field components 
$B_\varphi (r, \theta)$ (Eq.~A2), 
the radial component $B_r (r, \theta)$ (Eq.~A1),
the nonlinear parameter $\alpha (r, \theta)$ (Eq.~A4), and calculating
the derivative $\partial (B_\varphi \sin \theta) / \partial \theta $ 
yields then an expression where all terms cancel out,
$$
	[\nabla \times {\bf B}]_r = 0 \ ,
	\eqno(A15)
$$
as they cancelled out in the old approximation with $B_\theta=0$ also.

We list an overview of the accuracy orders of these approximations in 
Table 1.  In the overall, the lowest order among all components is 
of second order, in both the old and the new (VCA3-NLFFF) codes,
but the poloidal force-free component $[\nabla \times {\bf B}]_\theta$
improved from third-order to fifth-order, mostly due to the more
accurate approximation of $B_\theta \neq 0$ (A6).

\clearpage

\begin{table}
\center{
\tabletypesize{\normalsize}
\setlength{\tabcolsep}{0.05in}
\tabletypesize{\footnotesize}
\caption{Order of Accuracy in the Divergence-Freeness and Force-Freeness
Conditions in the Analytical Approximation of the (Old) VCA-NLFFF and
(New) VCA3-NLFFF Code.}
\medskip
\begin{tabular}{lll}
\hline
Equation			  & Old code	 	    & New code            \\
				  & VCA-NLFFF		    & VCA3-NLFFF	  \\
\hline
\hline
$ \nabla \cdot  {\bf B}$ 	  & $(b r \sin \theta)^2$ & $(b r \sin \theta)^2$\\
$[\nabla \times {\bf B}]_r$ 	  & $0$                   & $0$                 \\
$[\nabla \times {\bf B}]_\theta$  & $(b r \sin \theta)^3$ & $(b r \sin \theta)^5$\\
$[\nabla \times {\bf B}]_\varphi$ & $0$                   & $(b r \sin \theta)^3$\\
\hline
\end{tabular}
}
\end{table}

\begin{table}
\center{
\tabletypesize{\normalsize}
\setlength{\tabcolsep}{0.05in}
\tabletypesize{\footnotesize}
\caption{Energy parameters of the first 10 out of the 174 M- and X-class 
flare events in the longitude range of $[-45^\circ, +45^\circ]$, 
computed with the new VCA3-NLFFF Code. The full Table can be downloaded 
electronically from the journal as a machine-readable ASCII file.}
\medskip
\begin{tabular}{rrrrrrrrrrrrrrr}
\hline
 Nr & Date        & Time & GOES  & Helio  &   $E_p$ & $E_{np}$ & $E_{free}$ & $E_{diss}$ & $E_{nt,e}$ & $E_{nt,i}$ & $E_{dir}$ & $E_{th}$ & $E_{cme}$ & $E_{sum}$ \\
\hline
\hline
   3 & 2010-08-07 & 17:55 & M1.0 & N13E34 &   402.2 &   459.5 &    59.5 &    22.4 &     1.1 &     0.4 &     6.8 &     8.2 &     0.0 &     8.2 \\
   4 & 2010-10-16 & 19:07 & M2.9 & S18W26 &   164.6 &   173.6 &    12.7 &    11.7 &     1.4 &     0.5 &    17.4 &    19.2 &     0.1 &    19.3 \\
  10 & 2011-02-13 & 17:28 & M6.6 & S21E04 &   802.3 &   866.0 &    66.5 &    23.8 &     6.8 &     2.3 &    11.6 &    20.8 &     0.7 &    21.4 \\
  11 & 2011-02-14 & 17:20 & M2.2 & S20W07 &   998.6 &  1058.6 &    71.4 &    35.3 &     0.0 &     0.0 &    13.5 &    13.5 &     0.5 &    13.9 \\
  12 & 2011-02-15 & 01:44 & X2.2 & S21W12 &   966.8 &  1039.2 &    94.1 &    38.8 &    25.1 &     8.5 &    48.6 &    82.2 &     9.7 &    91.9 \\
  13 & 2011-02-16 & 01:32 & M1.0 & S22W27 &   959.2 &  1037.1 &   101.8 &    79.2 &     2.5 &     0.9 &     3.9 &     7.3 &     2.0 &     9.3 \\
  14 & 2011-02-16 & 07:35 & M1.1 & S23W30 &  1100.3 &  1209.6 &   125.0 &    59.8 &     0.0 &     0.0 &     4.4 &     4.4 &     0.0 &     4.4 \\
  15 & 2011-02-16 & 14:19 & M1.6 & S23W33 &  1054.5 &  1163.8 &   146.2 &    89.5 &     4.6 &     1.6 &     0.2 &     6.4 &     0.0 &     6.4 \\
  16 & 2011-02-18 & 09:55 & M6.6 & N15E05 &   727.3 &   779.1 &    87.3 &    83.1 &    11.8 &     4.0 &     0.1 &     4.2 &     0.0 &    15.9 \\
  17 & 2011-02-18 & 10:23 & M1.0 & N17E07 &   743.3 &   808.5 &    70.5 &     8.9 &     0.0 &     0.0 &     3.4 &     3.4 &     0.0 &     3.4 \\
\hline
\end{tabular}
}
\end{table}

\clearpage


\section*{REFERENCES} 

\def\ref#1{\par\noindent\hangindent1cm {#1}}

\ref{Ahmed, O.W., Qahwaji, R., Colak, T., et al. 
	2013, SoPh 283, 157}
\ref{Amari, T., Luciani, J.F., Aly, J.J., and Tagger, M. 
	1966, ApJ 466, L39}
\ref{Aschwanden, M.J., 
	2004, {\sl Physics of the Solar Corona. An Introduction},
        Praxis and Springer, Berlin, 216.}
\ref{Aschwanden, M.J. and Sandman, A.W. 
	2010, AJ 140, 723} 
\ref{Aschwanden, M.J.
	2012, A\&A 539, A2}
\ref{Aschwanden, M.J., Wuelser, J.P., Nitta, N.V., Lemen, J.R., Schrijver, C.J., et al.
 	2012, ApJ 756, 124} 
\ref{Aschwanden, M.J. 
	2013a, SoPh 287, 323} 
\ref{Aschwanden, M.J. 
 	2013b, SoPh 287, 369} 
\ref{Aschwanden, M.J. 
	2013c, ApJ 763, 115}  
\ref{Aschwanden, M.J. and Malanushenko, A.
 	2013, SoPh 287, 345} 
\ref{Aschwanden, M.J., Xu, Y., and Jing J. 
	2014a, ApJ 797:50, (Paper I)}
\ref{Aschwanden, M.J., Sun, X.D., and Liu,Y.
 	2014b, ApJ 785, 34} 
\ref{Aschwanden, M.J.
 	2015, ApJ 804, L20} 
\ref{Aschwanden, M.J., Boerner,P., Ryan, D., Caspi, A., McTiernan, J.M., and Warren, H.P. 
	2015a, ApJ 802, 53, (Paper II)}
\ref{Aschwanden, M.J., Schrijver, C.J., and Malanushenko, A.
 	2015b, SoPh 290, 2765} 
\ref{Aschwanden, M.J. 
	2016a, ApJ 831, 105, (Paper IV).}
\ref{Aschwanden, M.J.
 	2016b, ApJSS 224, 25} 
\ref{Aschwanden, M.J., O'Flannagain, A., Caspi, A., McTiernan, J.M., Holman, G., et al.
	2016a, ApJ 832, 27, (Paper III)}
\ref{Aschwanden, M.J., Crosby, N.B., Dimitropoulou, M., et al. 
	2016b, SSRv 198:47}
\ref{Aschwanden, M.J., Reardon, K., and Jess, D.
 	2016c, ApJ 826, 61} 
\ref{Aschwanden, M.J., Caspi, A., Cohen, C.M.S., Holman, G.D., Jing, J., et al.
	2017, ApJ 836, 17, (Paper V)}
\ref{Aschwanden, M.J. 
	2017, ApJ 847, 27, (Paper VI).}
\ref{Aschwanden, M.J., Gosic, M., Hurlburt, N.E., and Scullion, E.
 	2018, ApJ 866, 72} (VCA)
\ref{Aschwanden, M.J. and Gopalswamy,N. 
	2019, ApJ 877, 149, Paper VII} 
\ref{Aschwanden, M.J., Kontar, E.P., and Jeffrey, N.L.S. 
	2019, ApJ 881:1, (Paper VIII)}
\ref{Aschwanden, M.J. 
	2019a, {\sl New Millennium Solar Physics}, Springer Nature, 
	Switzerland, Science Library Vol. 458}
\ref{Aschwanden, M.J. 
	2019b, ApJ 874, 131} 
\ref{Bak, P., Tang, C., and Wiesenfeld, K. 
	1987, PhRvL 59/4, 381.}
\ref{Barnes, G., Leka, K.D., Schumer, E.A., et al. 
	2007, Space Weather 5/9, S09002}
\ref{Barnes, G. and Leka, K.D. 
	2008, ApJ 688, L107}
\ref{Barnes, G., Leka, K.D., Schrijver, C.J., et al. 
	2016, ApJ 829, 89}
\ref{B\'elanger, E., Vincent, A., and Charbonneau, P. 
	2007, SoPh 245, 141}
\ref{Bloomfield, D.S., Higgins, P.A., McAteer, R.T.J., et al. 
        2012, ApJ 747, L41}
\ref{Bobra, M.G. and Couvidat S. 
	2015, ApJ 798, 125}
\ref{Bobra, M.G. and Ilonidis, S. 
	2016, ApJ 821, 127}
\ref{Colak, T. and Qahwaji, R. 
	2009, Space Weather 7/6, S06001.}
\ref{Crosby, N.B., Aschwanden, M.J., and Dennis, B.R. 
	1993, SoPh 143, 275.}
\ref{Crosby, N.B., Vilmer, N., Lund, N., and Sunyaev, R. 
	1998, A\&A 334, 299}
\ref{Cui, Y., Li, R., Zhang, L. et al. 
	2006, SoPh 237, 45}
\ref{DeRosa, M.L., Schrijver, C.J., Barnes, G., Leka, K.D., Lites, B.W., Aschwanden, M.J., et.al. 
	2009, Apj 696, 1780}
\ref{Emslie, A.G., Dennis, B.R., Shih, A.Y., Chamberlin, P.C., Mewaldt, R.A., Moore, C.S. et al.
 	2012, ApJ 759, 71}
\ref{Falconer, D.A., Moore, R.L., and Gary, G.A. 
	2003, JGR 108, A10, 1380}
\ref{Falconer, D.A., Abdulnasser, F., Khazanov, I., et al. 
	2011, Space Weather 9/4, S04003}
\ref{Falconer, D.A., Moore, R.L., Barghouty, A.F., et al. 
	2012, ApJ 757, 32}
\ref{Georgoulis, M.K. and Rust, D.M. 
	2007, ApJ 661, L109}
\ref{Gibson, S.E. and Low, B.C. 
	2000, JGR 105/A8, 18187}
\ref{Gold, T. and Hoyle, F. 
	1960, MNRAS 120/2, 89.}
\ref{Hood, A.W. and Priest, E.R. 
	1979, SoPh 64, 303}
\ref{Jonas, E., Bobra, M., Shankar, V., et al. 
	2018, SoPh 293, 48} 
\ref{Kliem, B. and T\"or\"ok, T.
	2006, Phys.Rev.Let. 96/25, 255002}
\ref{Kliem, B., Lin, J., Forbes, T.G., Priest, E.R., and T\"or\"ok, T.
	2014, ApJ 789, 46}
\ref{Leka, K.D. and Barnes, G. 
	2003, ApJ 595, 1277}
\ref{Li, R., Wang, H.N., He, H. et al. 
        2007, Chinese J. Astron.Astrophys. 7/3, 441}
\ref{Lippiello, E., de Arcangelis, L., and Godano, C. 
	2010, A\&A 511, L2.}
\ref{Lu, E.T. 
	1995, ApJ 447, 416}
\ref{Mason, J.P. and Hoeksema, J.T. 
	2010, ApJ 723, 634}
\ref{Metcalf, T.R., Jiao, L., Uitenbroek, H., McClymont, A.N., and Canfield, R.C.
 	1995, ApJ 439, 474}
\ref{Nishizuka, N., Sugiura, K., Kubo, Y., et al. 
	2017, ApJ 835, 156}
\ref{Pesnell, W.D., Thompson, B.J., and Chamberlin, P.C. 
	2011, SoPh 275, 3.}
\ref{Priest, E.R.
 	2014, {\sl Magnetohydrodynamics of the Sun}, Cambridge: Cambridge University Press}
\ref{Rosner, R. and Vaiana, G.S. 
	1978, ApJ 222, 1104}
\ref{Sandman, A., Aschwanden, M.J., DeRosa, M., W\"ulser, J.P., Alexander, D. 
	2009, SoPh 259, 1.} 
\ref{Sandman, A.W., Aschwanden, M.J.  
	2011, SoPh 270, 503.} 
\ref{Schrijver, C.J., DeRosa, M.L., Title, A.M., and Metcalf,T.R.
 	2005, ApJ 628, 501}
\ref{Schrijver, C.J. 
	2007, ApJ 655, L117.}
\ref{Schrijver, C.J. 
	2016, ApJ 820, 103}
\ref{Strugarek, A. and Charbonneau, P.
 	2014, SoPh 289, 4137}
\ref{Sun, X., Hoeksema, J.T., Liu, Y., Kazachenko, M., and Chen, R.
	2017, ApJ 839:67}
\ref{Sun, X., Hoeksema, J.T., Liu, Y., Wiegelmann, T., Hayashi, K., Chen, Q., et al.
	2012, ApJ 748, 77}
\ref{Thalmann, J.K., Tiwari, S.K., and Wiegelmann, T. 
	2014, ApJ 780:102}
\ref{Thalmann, J.K., Linan, L., Pariat, E., and Valori, G. 
	2019, ApJ 880:L6}
\ref{T\"or\"ok, T., Kliem,B., and Titov,V.S.
 	2003, A\&A 413, L27}
\ref{Wang,W., Liu, R., Wang, Y., Hu, Q., Shen, C., Jiang, C., and Zhu, C.
 	2017, Nature Communications, 8,  1330}.
\ref{Warmuth, A., and Mann, G. 
	2016a, A\&A 588, A115}
\ref{Warmuth, A., and Mann, G. 
	2016b, A\&A 588, A116}
\ref{Warren, H.P., Crump, N.A., Ugarte-Urra,I., Sun,X., Aschwanden,M.J., and Wiegelmann,T.
 	2018, ApJ 860, 46} 
\ref{Wheatland, M.S., Sturrock, P.A., and McTiernan, J.M. 
	1998, ApJ 509, 448}
\ref{Wheatland, M.S.
        2000, ApJ 532, 1209.}
\ref{Wheatland, M.S.  Sturrock, P.A. and Roumeliotis, G. 2000, 
	ApJ, 540, 1150.}
\ref{Wiegelmann, T. 
	2004, SoPh 219, 87}
\ref{Wiegelmann, T., Inhester, B., and Sakurai, T. 
	2006, SoPh 233, 215.}
\ref{Wiegelmann, T., Thalmann, J.K., Inhester, B., Tadesse, T., Sun, X., Hoeksema, J.T. 
	2012, SoPh 281, 37}
\ref{Wiegelmann,T., Neukirch, T., Nickeler, D.H., et al. 
        2017, ApJSS 229, 18}
\ref{Zhu, X.S., Wang, H.N., Du, Z.L., et al. 
	2013, ApJ 768, 119}
\ref{Zhu, X.S. and Wiegelmann, T.,
        2018, ApJ 866, 130}

\clearpage


\begin{figure}
\centerline{\includegraphics[width=0.7\textwidth]{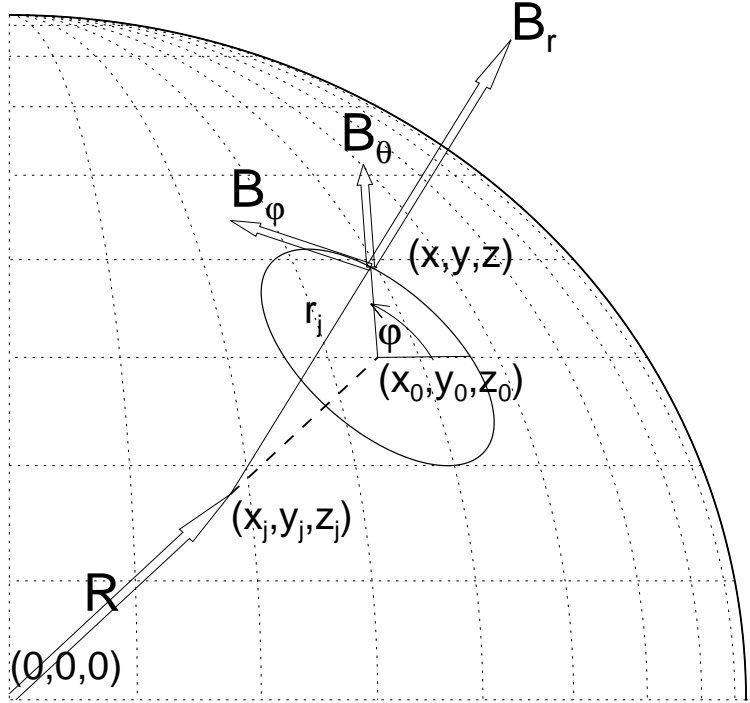}}
\caption{The geometry of the 3-D magnetic field components
${\rm B}=(B_r, B_{\varphi}, B_{\theta})$ in a coronal position
$(x,y,z)$ is shown, computed from a magnetic charge $j$ that is
buried at a subphotospheric position $(x_j,y_j,z_j)$ and is
helically twisted by an azimuth angle $\varphi$. The origin
$(0,0,0)$ of the spherical coordinate system is in the center of
the Sun. The central twist axis (dashed line) intersects an 
equi-potential surface at position $(x_0, y_0, z_0)$.
The radial field component $B_r$ points radially away from the
point charge at position $(x_j, y_j, z_j)$. The azimuthal 
magnetic field component $B_{\varphi}$ at location $(x,y,z)$ 
is orthogonal to the radial component $B_r$, and the poloidal
field component $B_{\theta}$ is orthogonal to both the
radial component $B_r$ and the azimuthal component $B_{\varphi}$.}
\end{figure}

\begin{figure}
\centerline{\includegraphics[width=0.8\textwidth]{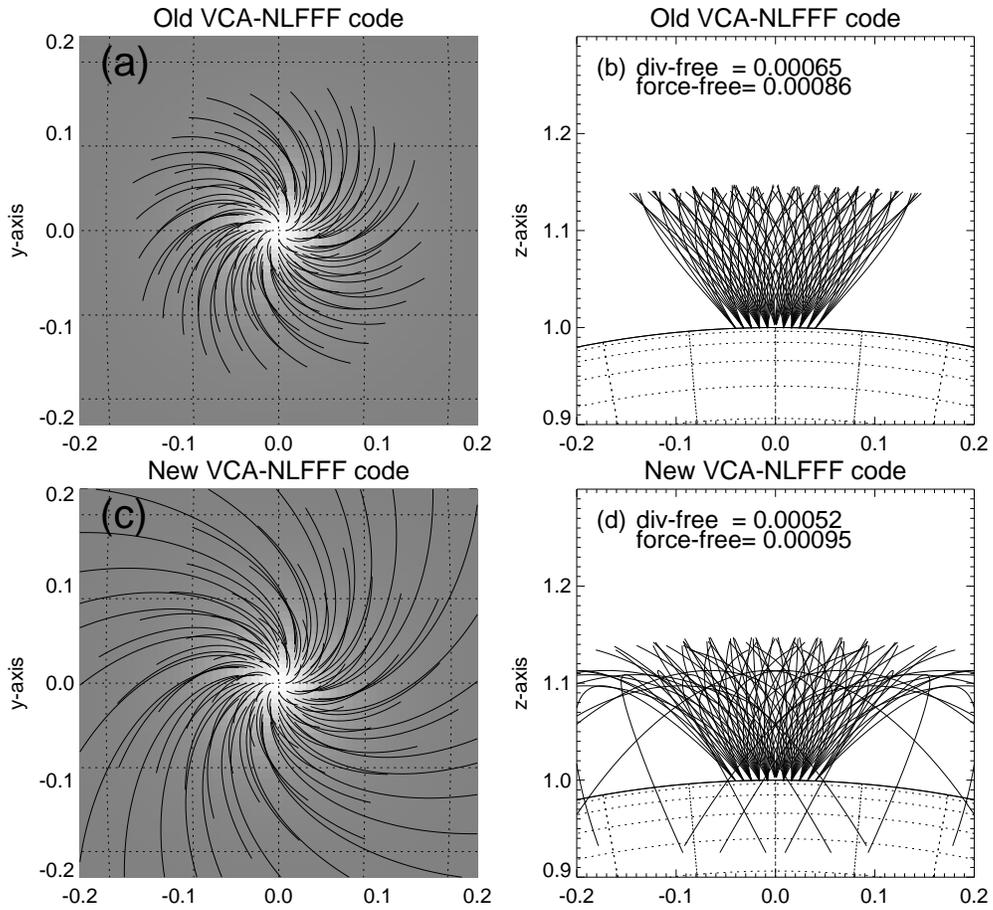}}
\caption{The magnetic non-potential field of a helically twisted
sunspot is computed with the old VCA-NLFFF code (top panels)
and with the new VCA3-NLFFF code (bottom panels), shown from a
top view (left panels) and a side view (right panels). Note that
the new field lines exhibit a curvature of the poloidal 
component $B_{\theta}$ (bottom right), which is neglected
in the old code (with $B_{\theta}=0$) (top right panel).}
\end{figure}

\begin{figure}
\centerline{\includegraphics[width=0.8\textwidth]{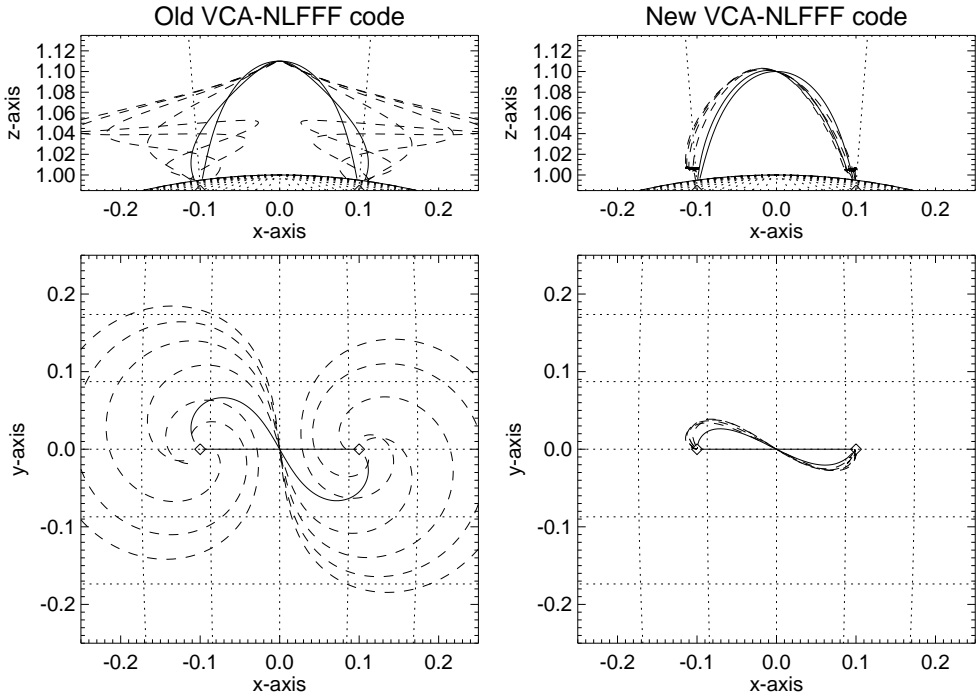}}
\caption{The non-potential magnetic field of a Gold-Hoyle
flux rope with 1 to 5 helical windings (dashed curves) and
the corresponding potential field solutions (solid) curves)
are shown for the old VCA-NLFFF code (left panels) and the
new VCA3-NLFFF code (right panels). Note that the neglect of
the poloidal component ($B_{\theta}=0$) leads to distorted
field lines in calculations with the old VCA-NLFFF code,
which are absent in the new VCA3-NLFFF code (right panels),
where force-free solutions are found always to have less
than one full turn of helical twist.}
\end{figure}

\begin{figure}
\centerline{\includegraphics[width=0.9\textwidth]{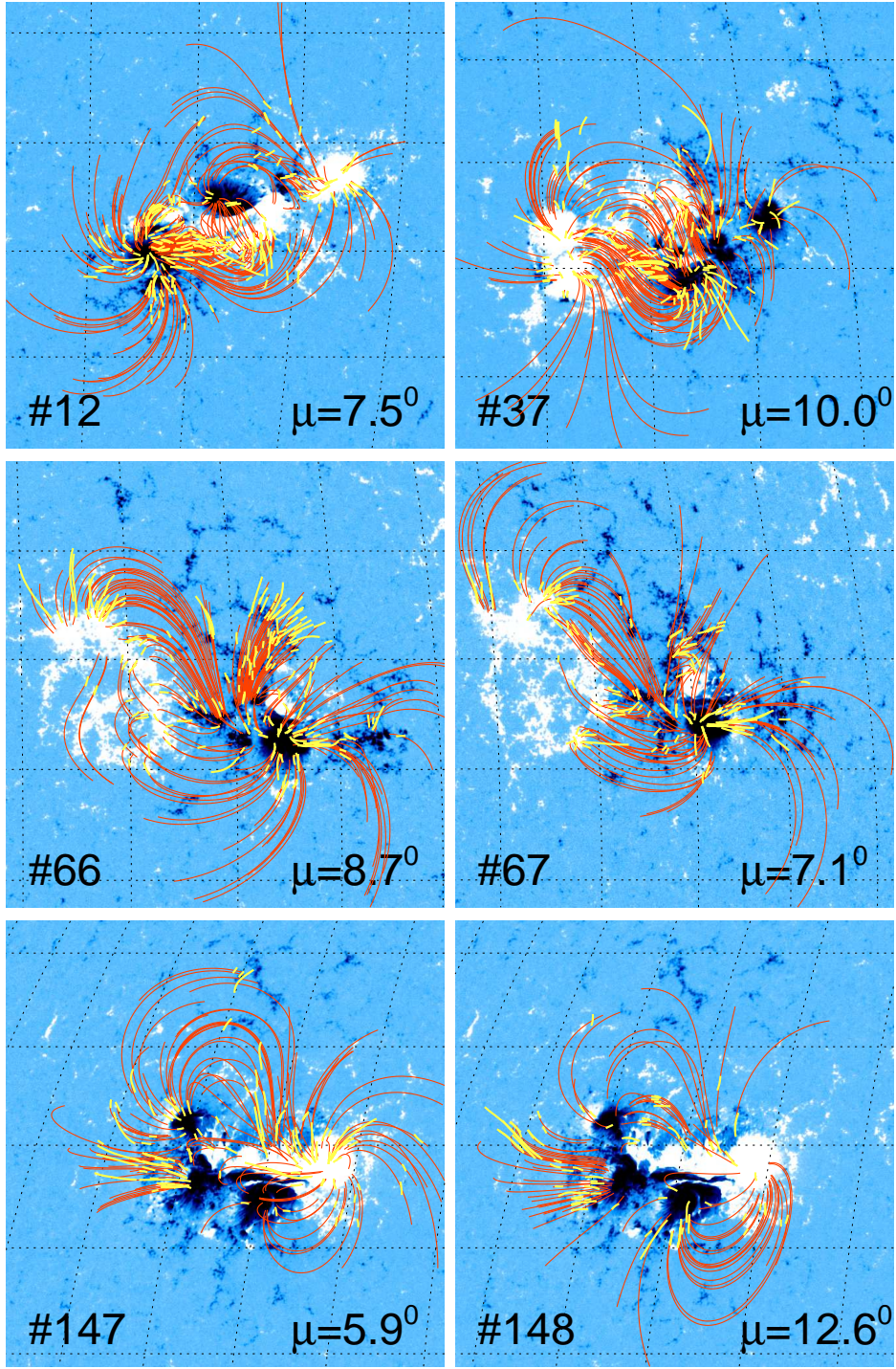}}
\caption{The HMI/SDO line-of-sight magnetogram (blue image)
with positive (white) and negative (black) magnetic polarities
is shown, along with automatically traced coronal loops
(yellow segments) and theoretical magnetic field lines
(red curves) that are computed with the VCA3-NLFFF code
and intersect the traced loop segments at their midpoint,
for 6 X-class flares observed with SDO. {\bf The misalignment
angle between the best-fit magnetic field model 
and observed loops is indicated with $\mu_2$.}}
\end{figure}

\begin{figure}
\centerline{\includegraphics[width=0.9\textwidth]{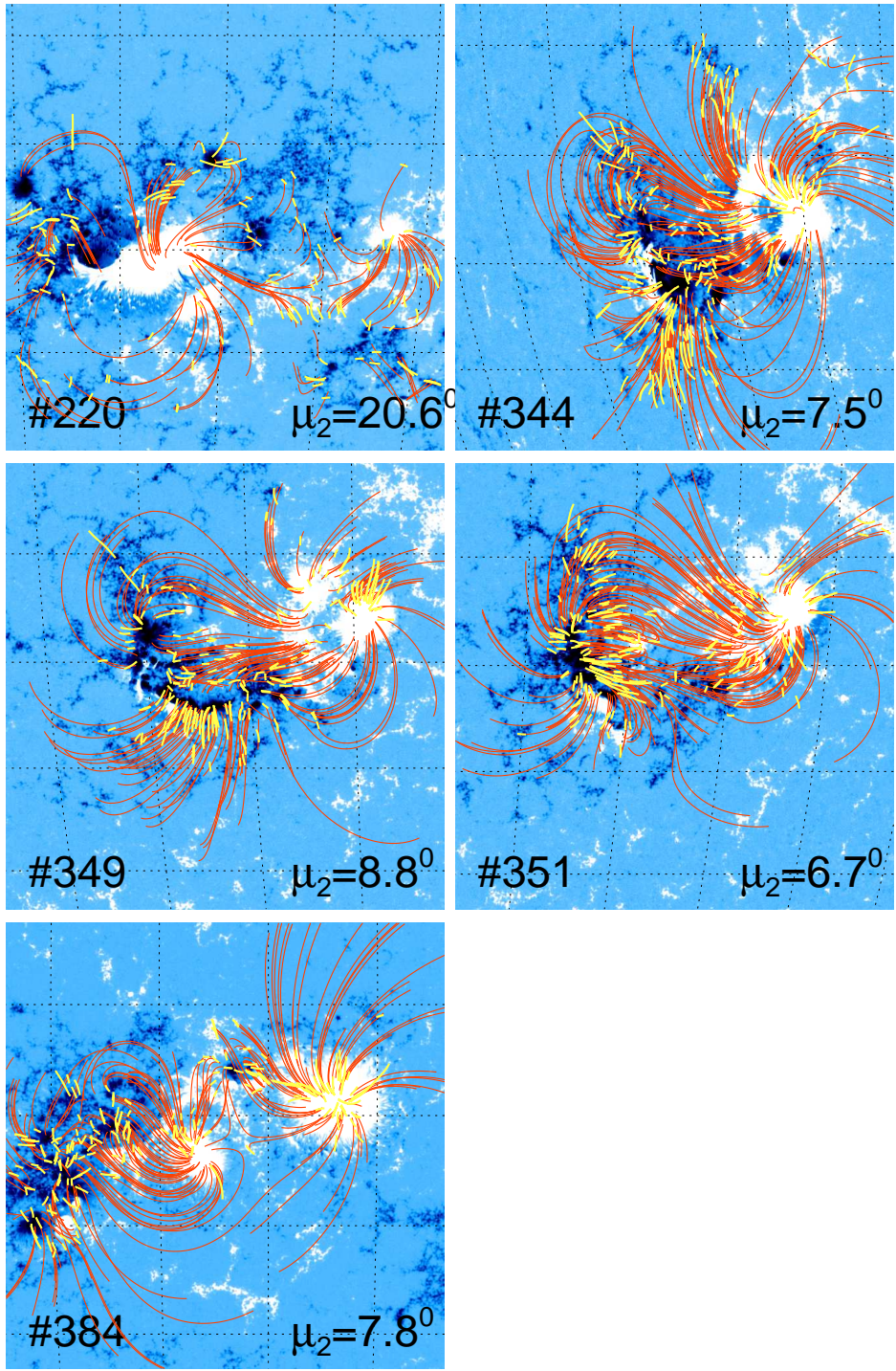}}
\caption{Similar representation as in Fig.~4, for 5 additional
X-class flares.}
\end{figure}

\begin{figure}
\centerline{\includegraphics[width=1.0\textwidth]{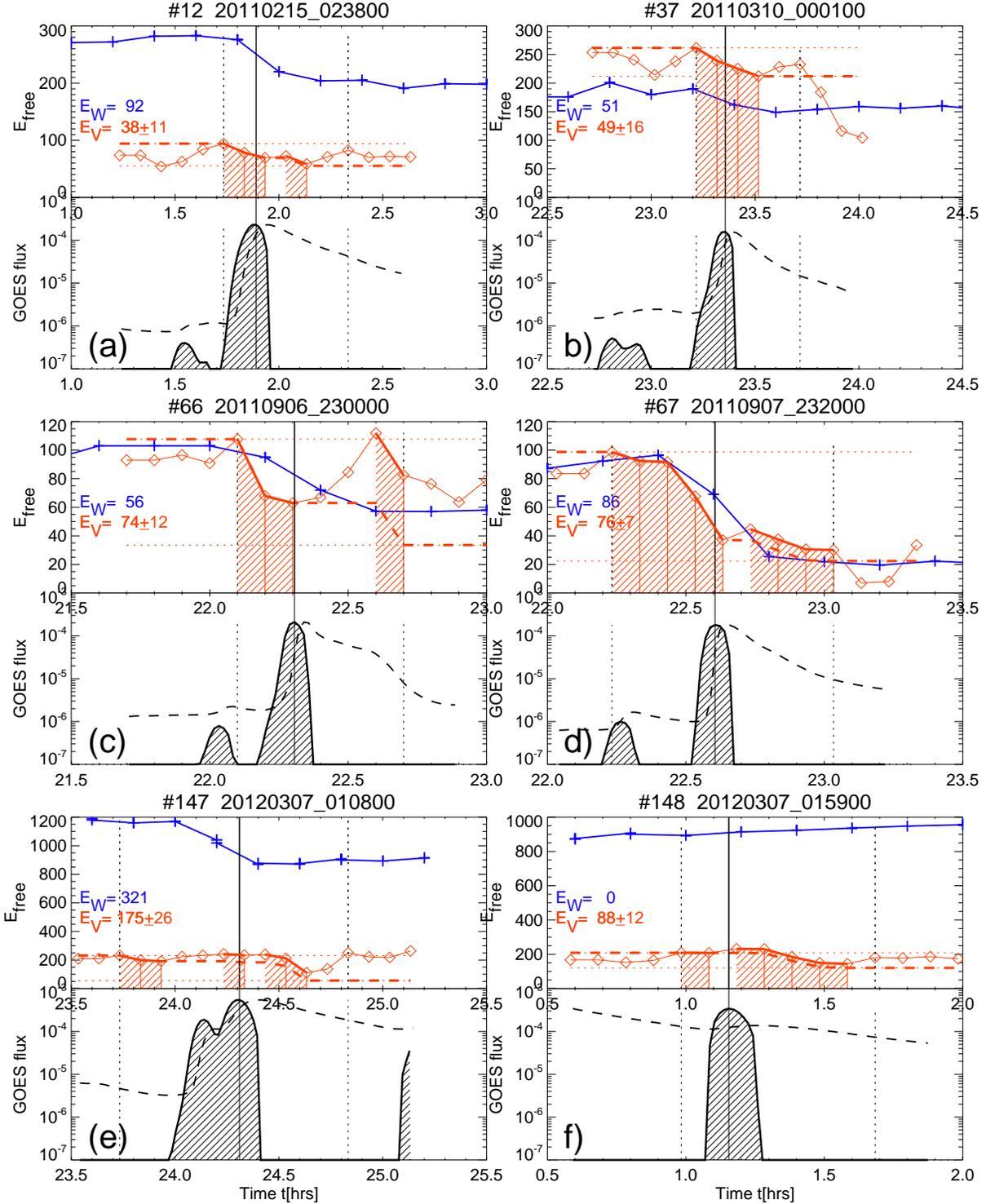}}
\caption{{\bf {\sl Top panels:} Time evolution of free energy 
for 6 X-class flares, computed with the VCA3-NLFFF code
(red), and compared with the Wiegelmann code (blue).
Additionally we show the cumulative energy decrease function 
(dashed red curve and hatched areas), the dissipated energy 
$E_V$ from the VCA3-NLFFF code, and $E_W$ from the Wiegelmann code,
the flare start and end time (vertical dotted lines), the
peak time of the GOES flux profile time derivative.
{\sl Bottom rows of panels:} The GOES flux profile (dashed curve)
and its time derivative (black hatched area).}}
\end{figure}

\begin{figure}
\centerline{\includegraphics[width=1.0\textwidth]{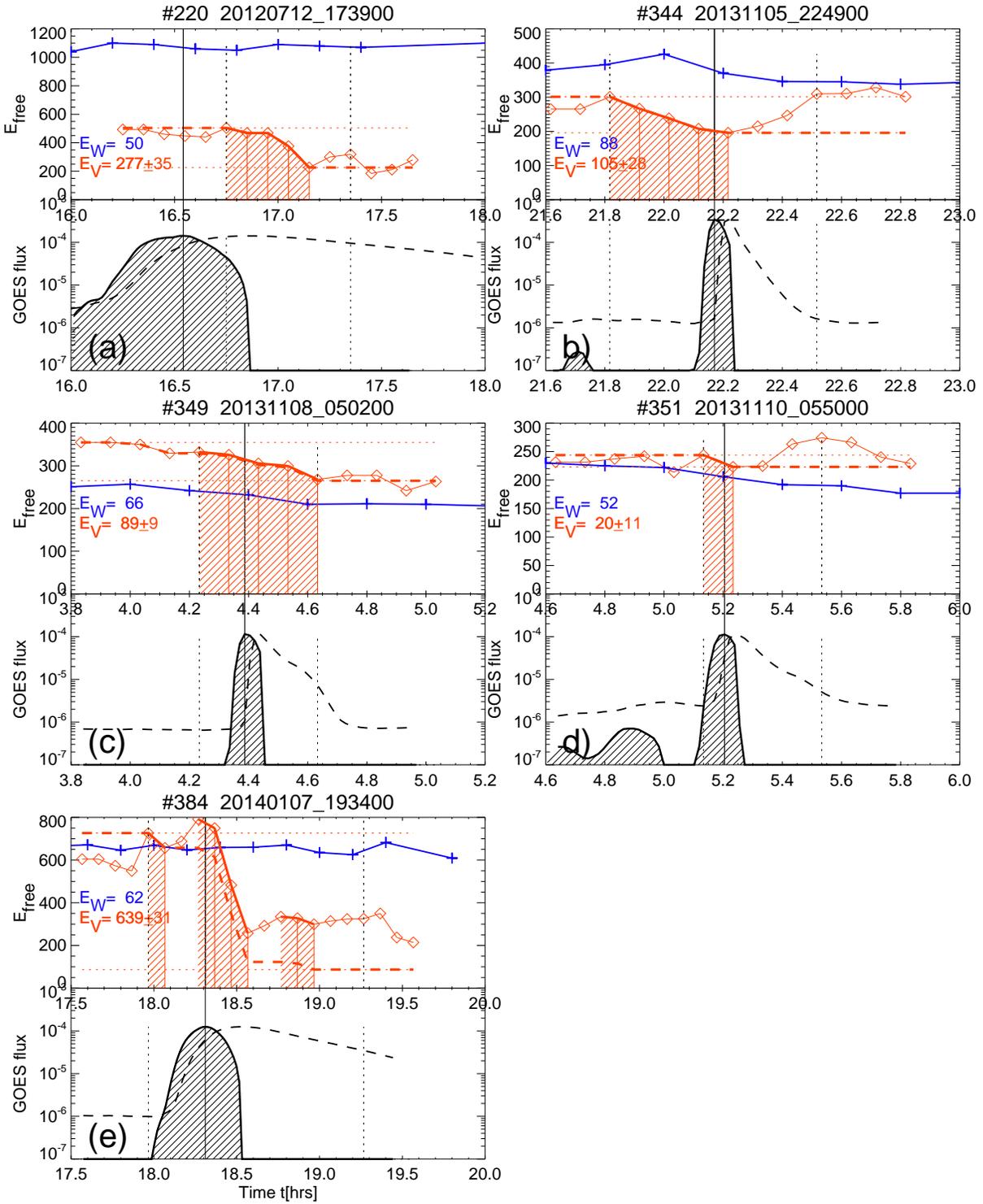}}
\caption{Similar representation as in Fig.~6, for 5 additional
X-class flare events.}
\end{figure}

\begin{figure}
\centerline{\includegraphics[width=1.0\textwidth]{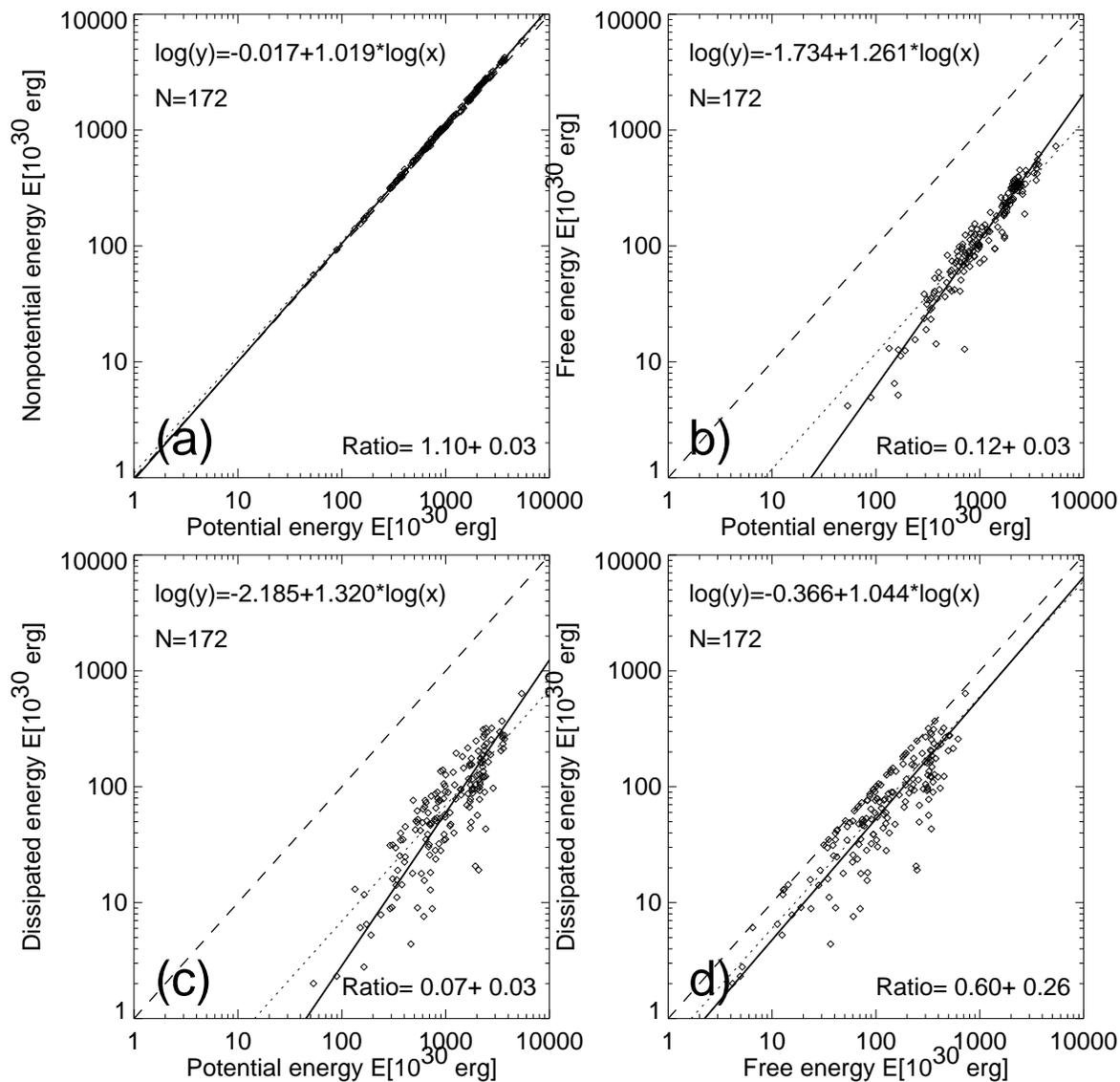}}
\caption{Scatterplots of potential, nonpotential, free, and
dissipated magnetic energies of our analyzed sample with
173 M- and X- GOES class flare events. Linear regression fits
in log-log space are indicated with thick black lines,
constant ratios with dotted lines, and equivalence with
dashed lines.}
\end{figure}

\begin{figure}
\centerline{\includegraphics[width=0.9\textwidth]{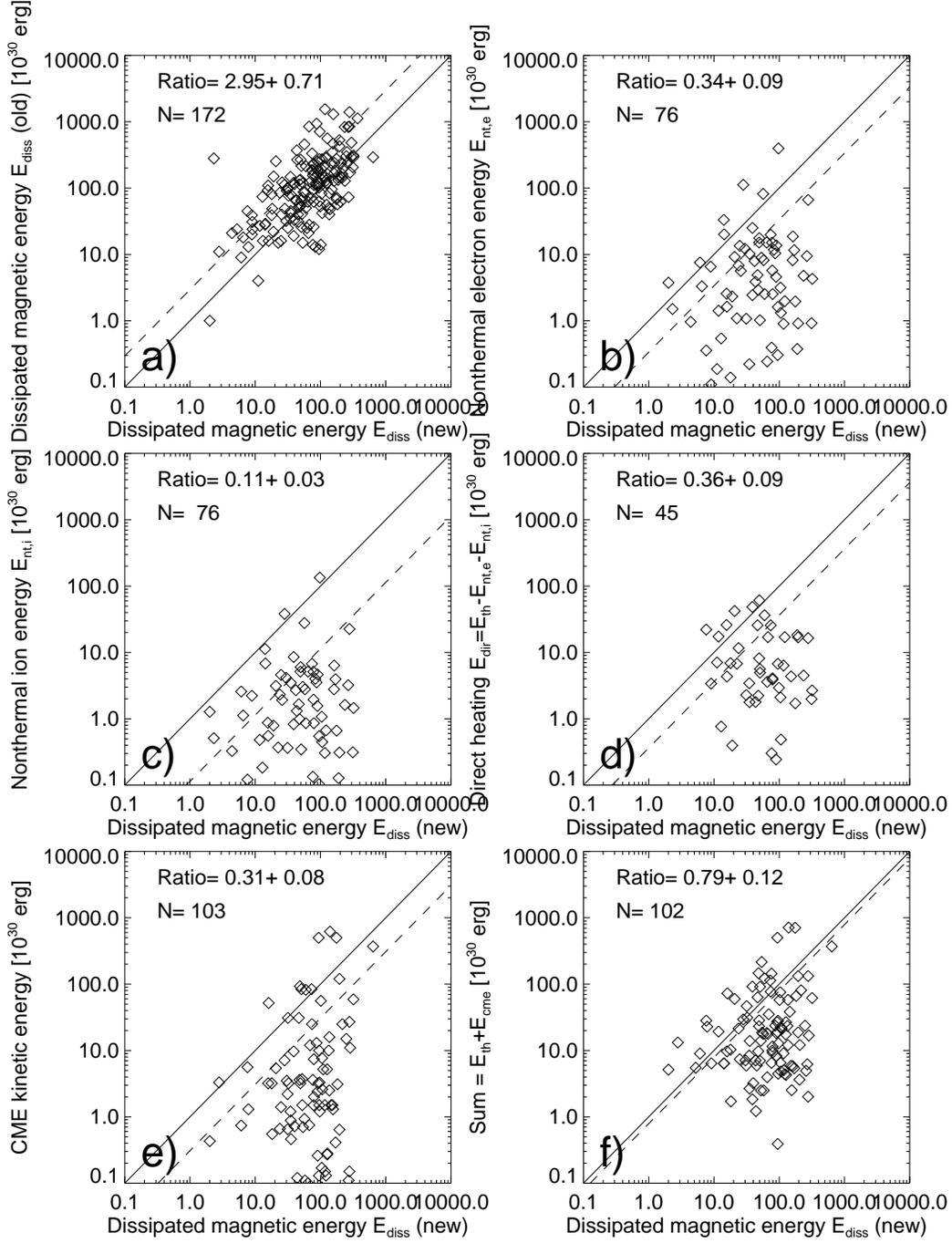}}
\caption{Scatterplot of the various forms of energies:
dissipated magnetic energies $E_{diss}^{old}$ (a), nonthermal
electron energies $E_{nt,e}$ (b), nonthermal energies in ions
$E_{nt,i}$ (c), direct heating $E_{dir}=E_{th}-E_{nt,e}-E_{nt,i}$ (d),
CME kinetic energy $E_{cme}$ (e), and the sum of primary energies
$E_{sum}=E_{nt,e}+E_{nt,i}+E_{dir}+E_{cme}=E_{th}+E_{cme}$ (f),
as a function of the dissipated magnetic energy 
$E_{diss}$ obtained with the new improved code.
Note that energy closure within $E_{sum}/E_{diss}=1.04\pm0.21$
is obtained for 44 events that have complete data (f).}
\end{figure}

\begin{figure}
\centerline{\includegraphics[width=1.0\textwidth]{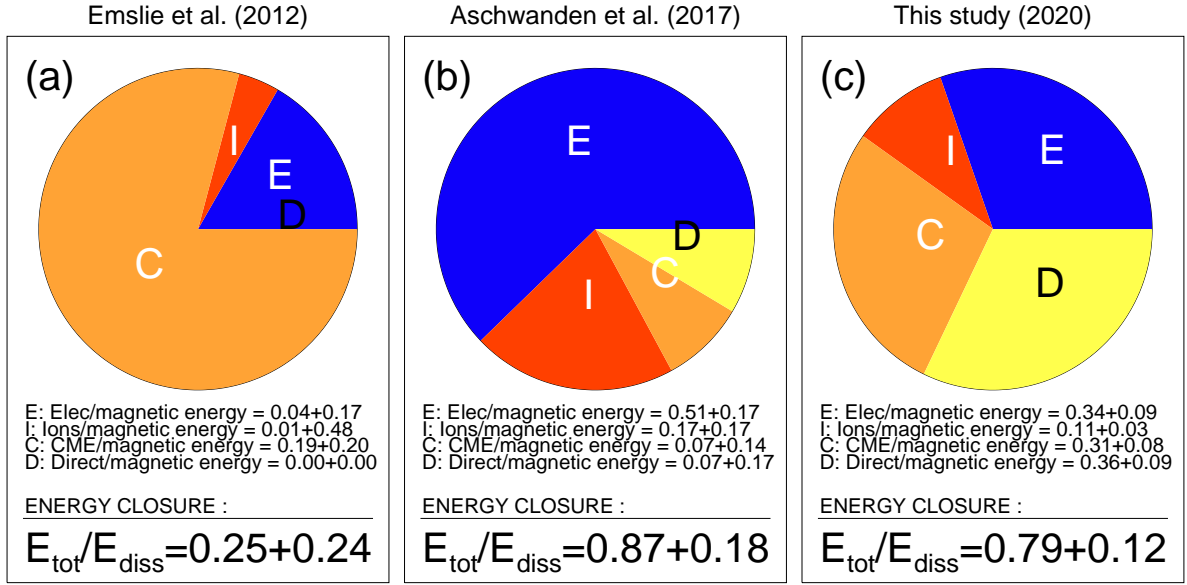}}
\caption{Occurrence frequency distributions of energy parameters
(histograms) with power law fits in log-log space (solid line).
The error of the slope $a$ is defined by $a/\sqrt{n}$, and the
total energy integrated over the entire distribution is indicated
with $E_{tot}$. Note that the slopes have typical values of
$a \approx 1.4-1.9$.}
\end{figure}

\begin{figure}
\centerline{\includegraphics[width=0.9\textwidth]{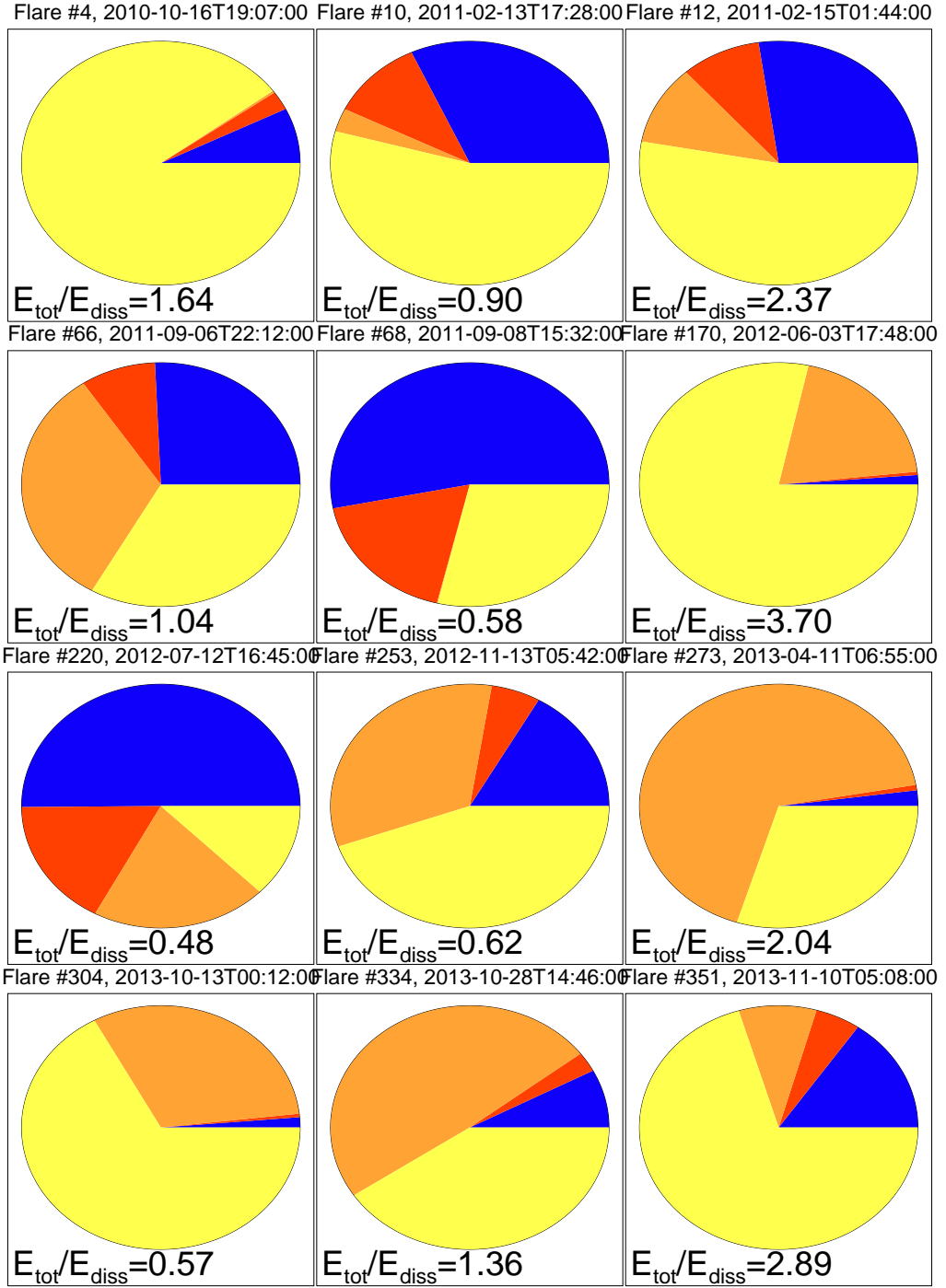}}
\caption{Pie chart diagrams the energy ratios of electrons
(blue), ions (red), CMEs (orange), and direct heating (yellow)
as a fraction of the total dissipated magnetic energy,
for the data set of Emslie et al.~(2012) (left panel), our earlier
work (Aschwanden et al.~2017) (middle panel), and the present study
(right panel).}
\end{figure}

\begin{figure}
\centerline{\includegraphics[width=0.9\textwidth]{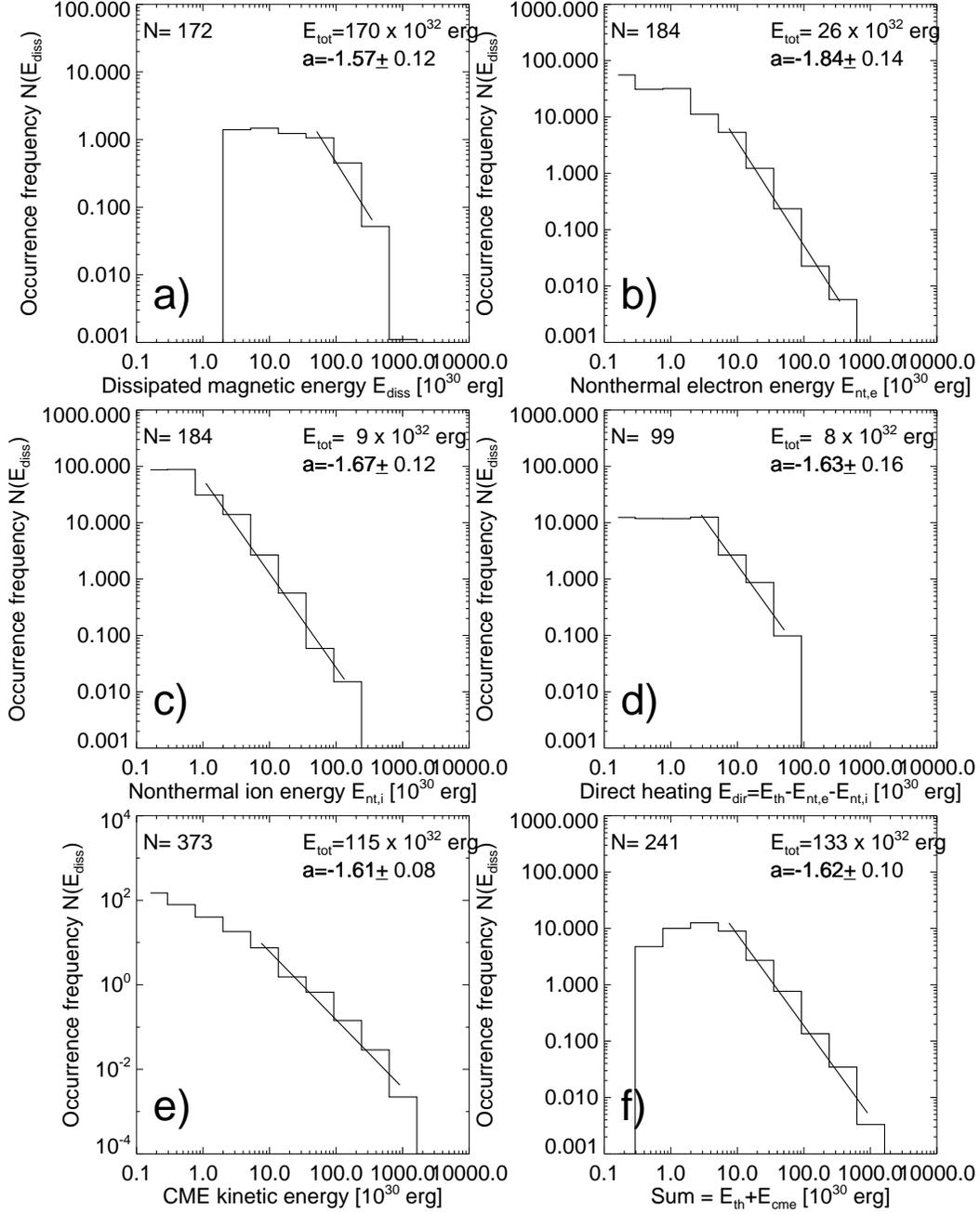}}
\caption{Pie chart diagrams the energy ratios of electrons
(blue), ions (red), CMEs (orange), and direct heating (yellow)
as relative fractions, shown for 12 flares. The ratios of the
total energy $E_{tot}=E_{th}+E_{cme}$ to the total dissipated
magnetic energy $E_{diss}$ is given in the lower left corners.}
\end{figure}

\begin{figure}
\centerline{\includegraphics[width=1.0\textwidth]{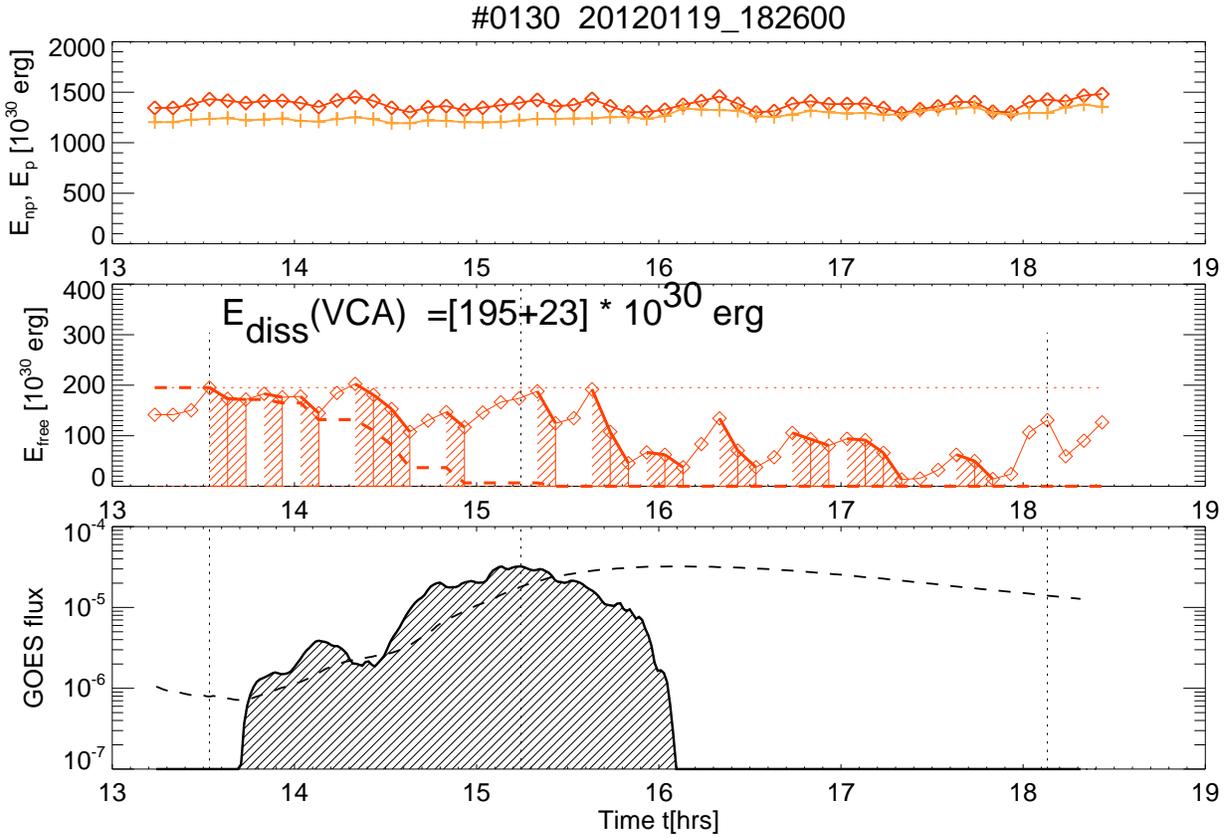}}
\caption{Time evolution of potential and non-potential energies
(top panel), the free energy (thick red curve in second panel),
time intervals of energy decreases (hatched red areas), 
cumulative free energy decrease function (dashed curve in 
middle panel), GOES 1-8 \ang\ flux (dashed curve in bottom
panel), and time derivative of GOES flux (hatched black areas
in bottom panel), for the long-duration flare \#130 on 
2012-Jan-19, 18:26 UT, which lasted over 4 hrs.}
\end{figure}

\end{document}